\documentclass[preprint2,tigten,letteredappendix,appendixfloats]{aastex631}

\usepackage{hyperref}
\usepackage{amsmath}

\begin{document}

\title{A Strongly Lensed Dusty Starburst of an Intrinsic Disk Morphology at Photometric Redshift of $z_{\rm ph}>7$}

\author[0000-0003-4952-3008]{Chenxiaoji Ling}
\affiliation{National Astronomical Observatories, Chinese Academy of Science, Beijing, 100101, China}

\author[0000-0001-7957-6202]{Bangzheng Sun}
\affiliation{Department of Physics and Astronomy, University of Missouri, Columbia, MO\,65211, USA}

\author[0000-0003-0202-0534]{Cheng Cheng}
\affiliation{Chinese Academy of Sciences South America Center for Astronomy, National Astronomical Observatories, CAS, Beijing, 100101, China}

\author[0000-0001-6800-7389]{Nan Li}
\affiliation{National Astronomical Observatories, Chinese Academy of Science, Beijing, 100101, China}
\affiliation{School of Astronomy and Space Science, University of Chinese Academy of Science, Beijing 100049, China}

\author[0000-0003-3270-6844]{Zhiyuan Ma}
\affiliation{Department of Astronomy, University of Massachusetts, Amherst, MA\, 01003, USA}

\author[0000-0001-7592-7714]{Haojing Yan}
\affiliation{Department of Physics and Astronomy, University of Missouri, Columbia, MO\,65211, USA}

\correspondingauthor{Haojing Yan}

\begin{abstract}

    We present COSBO-7, a strong millimeter (mm) source known for more than 
sixteen years but was just revealed its near-to-mid-IR counterpart by the James 
Webb Space Telescope (JWST). The precise pin-pointing by the Atacama Large 
Millimeter Array (ALMA) on the exquisite NIRCam and MIRI images show that it is 
a background source gravitationally lensed by a single foreground galaxy, and 
the analysis of its spectral energy distribution by different tools is in favor 
of photometric redshift at $z_{\rm ph}>7$. Strikingly, our lens modeling based
on the JWST data shows that it has a regular, disk morphology in the source 
plane. The dusty region giving rise to the far-IR-to-mm emission seems to be 
confined to a limited region to one side of the disk and has a high dust 
temperature of $>90$~K. The galaxy is experiencing starburst both within and 
outside of this dusty region. After taking the lensing magnification of 
$\mu\approx 2.5$--3.6 into account, the intrinsic star formation rate is 
several hundred $M_\odot$~yr$^{-1}$ both within the dusty region and across the
more extended stellar disk, and the latter already has $>10^{10}M_\odot$ of 
stars in place. If it is indeed at $z>7$, COSBO-7 presents an extraordinary 
case that is against the common wisdom about galaxy formation in the early 
universe; simply put, its existence poses a critical question to be answered:
how could a massive disk galaxy come into being so early in the universe and
sustain its regular morphology in the middle of an enormous starburst?  

\end{abstract}


\section{Introduction}

   Systematic investigation of dust-embedded star formation in external 
galaxies started from the mid-to-far-IR survey carried out by the InfraRed 
Astronomy Satellite four decades ago
\citep[see e.g.,][]{deJong1984, Sofier1984, Lonsdale1984}. The climax of 
these early studies was the discovery of Ultra-luminous Infrared Galaxies in 
the local universe
\citep[ULIRGs;][]{Houck1984, Houck1985, Aaronson1984}, which have enormous IR
luminosity of $L_{\rm IR}\geq 10^{12} L_\odot$ (integrated over the rest-frame
8--1000~$\mu$m) but are faint in optical.
In the late 1990s to early 2000s, a 
series of new instruments opened up the submillimeter (submm) and millimeter 
(mm) window from the ground and detected a new population of galaxies that were 
collectively called ``submillimeter galaxies'' (SMGs; see \citet[][]{Blain2002} 
for an early review). Through the pin-pointing of their positions by radio 
interferometry and the subsequent spectroscopy of their optical counterparts
\citep[see e.g.,][]{Barger2000, Chapman2003}, a consensus has been reached that
SMGs are galaxies mostly at $z\approx 2$--3 with ULIRG-like IR luminosity, and 
their continuum submm/mm emissions, which must be due to heated dust, are 
attributed to extreme star formation, often with star formation rates (SFRs) 
$>100$~$M_\odot$~yr$^{-1}$.

    The past decade has witnessed great advancements in the study of
high-redshift (high-$z$) dusty galaxies. Numerous far-IR galaxies (FIRGs), 
which are akin to SMGs, have been cataloged by the wide-field far-IR surveys of 
the Herschel Space Observatory. The unprecedented sensitivity of the
Atacama Large Millimeter Array (ALMA) has allowed the precise localization of
SMGs and FIRGs in significant numbers to enable statistical studies at various
wavelengths. A more general term, ``dusty star forming galaxies''
\citep*[DSFGs;][]{Casey2014}, has become widely in use to refer to galaxies
beyond the local universe that are detected in the far-IR-to-mm wavelengths,
with the implicit understanding that their continuum emission in this regime 
is due to dust heated by star formation. While still subject to some debate, 
it is recognized that DSFGs, once a population of galaxies discovered in a
special wavelength range, are not necessarily special as compared to the 
general population of star-forming galaxies in terms of their global 
properties; for example, a large fraction of DSFGs seem to follow the tight
SFR versus stellar mass relation (the so-called ``main sequence'') of 
star-forming galaxies \citep[e.g.,][]{Michalowski2012, daCunha2015}. In this
new context, any galaxies with continuum dust emission signifying their
dust-embedded star formation can be categorized as DSFGs, including some 
high-$z$ Lyman-break galaxies (LBGs) \citep[e.g.,][]{Capak2015, Bethermin2020}.
Somewhat surprisingly, continuum dust emissions have been detected in a few 
quasars and LBGs deep in the epoch of reionization (EoR), the earliest of which 
are at $z>7$ 
\citep[e.g.,][]{Venemans2012, Watson2015, Hashimoto2019a, Bakx2020, Sommovigo2022}. 
This implies that at least some galaxies at even earlier epochs must already be 
forming a substantial amount of dust. Dusty starbursts, the most extreme ones 
among DSFGs, have also been seen in the EoR 
\citep[][]{Riechers2013, Strandet2017}. This raises an interesting question:
are there starbursts in the EoR that are
\emph{not} enshrouded by dust? While there is no obvious reason that they 
should not exist, no such objects have been found in surveys for LBGs despite
that they would be very easy to detect because of their extreme brightness in
the rest-frame UV.

    The advent of James Webb Space Telescope (JWST) is bringing the study of
high-$z$ DSFGs to a new level. In merely over a year, the synergy of JWST and
ALMA has brought us a lot of details on the connection between dust-obscured and
unobscured (``exposed'') stellar populations in DSFGs
\citep[e.g.,][]{Cheng2022, Chen2022, Cheng2023, Huang2023, Tadaki2023, Liu2023, 
Barger2023, Kamieneski2023, Fujimoto2023, Hashimoto2023, Boogaard2023, 
Yoon2023, Rujopakarn2023, AM2023, Sun2024, Killi2024},
including a subset among the so-called ``HST-dark'' galaxies that are 
faint or even invisible at $\lambda\lesssim 1.6$~$\mu$m 
\citep[e.g.,][]{Zavala2023,Kokorev2023,Smail2023}.
Here we report on a peculiar dusty starburst that could possibly change 
the existing view of DSFG formation in the EoR. This object is known as a 
bright SMG for more than sixteen years, however the ALMA pinpointing of its
location on the existing Hubble Space Telescope (HST) images would only
associate it with a bright spiral galaxy at $z=0.359$. Its true counterpart in 
the near-to-mid-IR wavelengths was only recently revealed in the JWST NIRCam
and MIRI data. Using the MIRI data alone, \citet[][]{Pearson2024} point out
that this source is strongly lensed by the foreground galaxy at $z=0.359$.
While many strongly lensed DSFGs have been seen among bright SMGs, we find this
one special. Based on our comprehensive analysis of both the NIRCam and the
MIRI data, its photometric redshift is likely at $z_{\rm ph}> 7$ and its
reconstructed image in the source plane has a disk morphology.
A dusty starburst hosted by a disk galaxy at such a high redshift, 
if confirmed, will not only be a record-breaker but also pose a severe 
challenge to our current theories of galaxy formation in the early universe. 

   This paper presents the results from our study of this object,
organized as follows. The relevant multi-wavelength data are described in 
Section 2, and the photometry is given in Section 3. The analysis of the
spectral energy distributions (SEDs) by four different tools is given in 
Section 4. The lens model and image reconstruction are detailed in Section 5. We 
discuss our results in Section 6 and conclude with a summary in Section 7. All 
magnitudes are in the AB system. We adopt the following cosmological parameters 
throughout: 
$H_0=70$~km~s$^{-1}$~Mpc$^{-1}$, $\Omega_M=0.3$ and $\Omega_\Lambda=0.7$.

\section{Data and Source Characterization}

\subsection{The SMG and its precise location}



Our source was first discovered as a bright mm object in the COSBO survey 
\citep[source name ``MM J100024$+$021748'';][]{Bertoldi2007}, which was a 
1.2~mm imaging survey in the COSMOS field done by the Max-Planck Millimeter 
Bolometer Array (MAMBO-2) at the 30-m IRAM telescope. For simplicity, hereafter 
we refer to it as ``COSBO-7'' according to its ID in the COSBO catalog. This 
source was later confirmed at different mm and sub-mm wavelengths, and these 
measurements are summarized in Table~\ref{tab:literature}.
 
  

The sub-arcsec position of this source has been located by the ALMA Band 7 
(870~$\mu$m) observations made on May 15, 2018 (beam size $\sim$0\farcs75; 
program ID 2016.1.00463.S, PI: Y. Matsuda), which was reported in 
\citet[][]{Simpson2020} as ``AS2COSMOS0005.1'' at 
R.A. = 10:00:23.97, Decl. = 02:17:50.1 (J2000.0). We adopt the position in the 
publicly available catalog (version 20220606) from the A$^3$COSMOS project
\citep[][]{Liu2019}:
R.A. = 10:00:23.98, Decl. = 02:17:49.99 (J2000.0). Both are consistent 
with the indirect pinpointing of the VLA 3~GHz observations at a similar 
spatial resolution 
\citep[][beam size $\sim$0\farcs75]{Smolcic2017}, which gives 
R.A. = 10:00:23.95, Decl. = 02:17:50.03 (J2000.0).

\begin{deluxetable*}{llllll}
    \tablecaption{Observations of COSBO-7 in the literature}
    \label{tab:literature}
    \tablehead{
        Reference         & Survey/Instrument & Band                     & R.A.          & Decl.                        & Flux density   }
    \startdata
        \citet[][]{Bertoldi2007} & COSBO/MAMBO-2      & 1.2 mm                   & 10:00:24    & 02:17:48         & 5.0       $\pm$\ 0.9  mJy    \\ 
    \citet[][]{Aretxaga2011} & AzTEC      & 1.1 mm                    & 10:00:24.2  & 02:17:48.7    & 3.1 $\pm$~1.2 mJy   \\ 
    \citet[][]{Casey2013}    & SCUBA-2     & 450 $\mu$m & 10:00:23.8  & 02:17:51   & 12.7 $\pm$ 5.42 mJy \\
    \citet{Casey2013}    & SCUBA-2     & 850 $\mu$m & 10:00:23.8  & 02:17:51   &  8.42    $\pm$ 0.92 mJy \\ 
    \citet{Geach2017}        & S2CLS      & 850 $\mu$m  & 10:00:24.00 & 02:17:50.6 &  9.25       $\pm$ 1.14 mJy \\ 
    \citet{Simpson2019}  & S2COSMOS   & 850 $\mu$m  & 10:00:23.93 & 02:17:51.6 & 10.3  $\pm$ 0.8 mJy  \\ 
    \citet{Smolcic2017} & VLA-COSMOS & 3 GHz & 10:00:23.95 & 02:17:50.03 & 28.6     $\pm$ 2.8 $\mu$Jy \\ 
    \citet{Schinnerer2010} & VLA-COSMOS & 1.4 GHz &10:00:23.99  & 02:17:49.96 &  187         $\pm$ 12 $\mu$Jy \\ 
    \citet{Liu2019}$^\dagger$  & A$^3$COSMOS & 870 $\mu$m & 10:00:23.98 & 02:17:49.99 & 10.446         $\pm$ 0.601 mJy \\   
    \enddata
    \tablenotetext{}{Notes. $^\dagger$ See also \citet[][]{Simpson2020}.}.
\end{deluxetable*}

\subsection{ALMA mm/sub-mm data and Herschel far-IR data}

   In addition to the aforementioned ALMA Band 7 data, this source also has 
Band 4 (2.07~mm) data taken on September 19, 2022 (program ID 2021.1.00705.S, 
PI: O. Cooper) that are now public. For the purpose of this work, we reduced 
all the public ALMA data in a uniform way. We utilized the default pipeline
{\sc ScriptForPi.py} and performed {\sc tclean} using the Common Astronomy 
Software Applications \citep[CASA;][]{2022PASP..134k4501C} to obtain the 
continuum image in each band. The cleaning was done to 3~$\sigma$ with the 
default natural weighting parameters of WEIGHTING = ``BRIGGS'' and 
ROBUST = 2.0. The final maps reach the root mean square (rms) of 0.180 and 
0.075~mJy~beam$^{-1}$ in Band 7 and 4, respectively, and have the beams of 
(bmaj, bmin, PA) = (0\farcs78, 0\farcs72, $-$85.9$^{\rm o}$)
and (1\farcs56, 1\farcs33, $-$63.0$^{\rm o}$) in these two bands, 
respectively, where bmaj, bmin and PA are the major axis, minor axis and 
positional angle, respectively.

   The COSMOS field was observed by the Herschel Multi-tiered Extragalactic
Survey \citep[HerMES;][]{Oliver2012} using the SPIRE imager in 250, 350 and 
500~$\mu$m, which had beam sizes of $\sim$18\arcsec, 25\arcsec~and 36\arcsec, 
respectively. We make use of the maps included in its fourth data release,
which have the pixel scales of 6\farcs0, 8\farcs3~and 12\farcs0 in these three
bands, respectively.

\subsection{JWST near-to-mid IR Data}

\begin{figure*}
    \centering
    \includegraphics[width=\textwidth,height=\textheight,keepaspectratio]{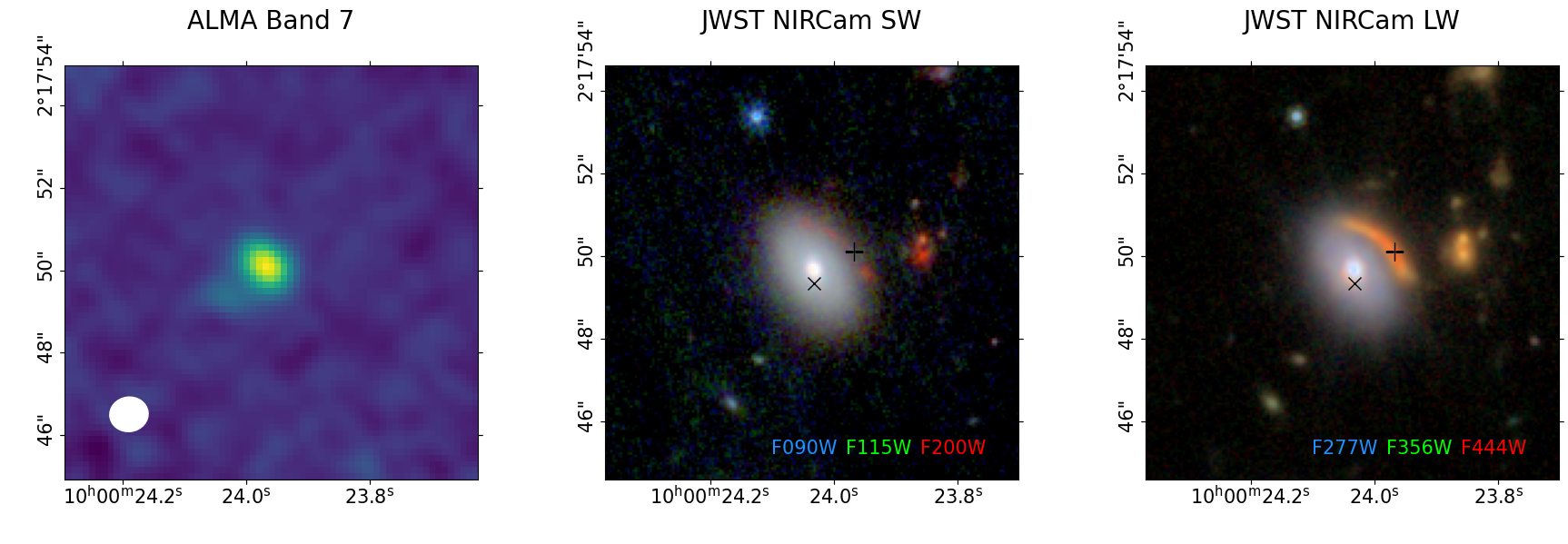}
    \caption{Images of COSBO-7 and its vicinity, with the coordinates labeled.
From left to right, these are in ALMA Band 7 (870~$\mu$m) and JWST NIRCam SW 
and LW channels, respectively. The white ellipse in the left panel indicates 
the beam size in 870~$\mu$m. The bright 870~$\mu$m source has a close, fainter 
companion in its southeastern direction, and their centers are only 1\farcs25 
apart. The NIRCam images are color-composites made of F090W (blue), F115W 
(green) and F200W (red) in SW and F277W (blue), F356W (green) and F444W (red) 
in LW, respectively. The positions of the bright 870~$\mu$m source and its weak 
companion are labeled as the black ``+'' and ``x'' symbols in these two images, 
respectively. Apparently, this is a strong gravitational lensing system: the 
NIRCam counterpart of the bright 870~$\mu$m source is the very red arc around 
the bright disk galaxy at $z=0.359$, which must be an image of a background 
source produced by this disk galaxy as the lens. The weak 870~$\mu$m source is 
likely the counter image of the brighter one, however its NIRCam counterpart is 
swamped by the lens galaxy.
}
    \label{fig:the_arc}
\end{figure*} 

The JWST NIRCam and MIRI imaging data of this source are from the Public 
Release IMaging for Extragalactic Research (PRIMER) program in Cycle 1 
\citep[][]{Dunlop2021}. The NIRCam observations were done in eight bands: 
F090W, F115W, F150W, F200W in the short-wavelength (SW) channel and F277W, 
F356W, F410M, and F444W in the long-wavelength (LW) channel, respectively, and 
the MIRI observations were done in F770W and F1800W.

We reduced these data on our own. We started from the Stage 1 ``uncal.fits'' 
products retrieved from the Mikulski Archive for Space Telescopes (MAST), which 
are single exposures after the ``Level 1b'' processing by the default JWST 
calibration procedures. Our further process used the JWST pipeline version 
1.10.2 in the context of {\tt jwst\_1089.pmap} for the NIRCam data and version 
1.12.5 in the context of {\tt jwst\_1183.pmap} for the MIRI data. 
We first ran the Stage 1 of the calibration pipeline, {\tt calwebb\_detector1}, 
which applies the detector-level corrections to the ``uncal.fits'' images. 
The outputs were then processed through Stage 2, {\tt calwebb\_image2}, 
which mainly applies the flat-field correction and the flux calibration.
We adopted most of the default parameters but with a few changes: 
(1) for the NIRCam images, we expanded the large ``jump'' events, 
also known as the ``snowball'', by 1.5 times the radius of the event to the 
neighboring pixels for better masking; 
(2) for the MIRI images, we enabled the detection and masking of ``showers''; 
(3) we removed the ``1/f'' readout noise patterns in the NIRCam SW images by 
using an external recipe ``image1overf''
\footnote{\href{https://github.com/chriswillott/jwst}{https://github.com/chriswillott/jwst}};
(4) we removed the vertical and/or horizontal stripe-like noise patters in the 
MIRI images following the recipe by \cite{Yang2023}; 
(5) we used the external tool ``JHAT''
\footnote{\href{https://github.com/arminrest/jhat}{https://github.com/arminrest/jhat}} 
to align each single exposure to the astrometric grid defined by the HST
images from the CANDELS program \citep[][]{Grogin2011,Koekemoer2011}; 
and (6) we estimated and subtracted a smooth, 2-dimensional background from all single exposures before stacking. For the MIRI images, we also masked the
entire coronagraph regions.
After all these processes, the 
individual exposures in each band were then stacked into one mosaic with a 
pixel scale of 0\farcs06. 
The images are in the unit of MJy~sr$^{-1}$, 
which translates to a magnitude zero-point of 26.581. 

\subsection{COSBO-7: a strongly lensed system due to a single foreground galaxy}

Figure \ref{fig:the_arc} shows the ALMA 870~$\mu$m image of COSBO-7 and its 
color-composite images in the NIRCam SW and LW. Based on the ALMA position 
(``+'' symbol), it is obvious that COSBO-7 corresponds to the red arc in the 
NIRCam LW bands, which must be an image produced by the strong gravitational 
lensing due to a single disk galaxy in the foreground. This foreground galaxy 
has the spectroscopic redshift of $z=0.359$ \citep[][]{Hasinger2018}. 

There are two notable points regarding this lensing system. First, the arc is 
extremely red: while being very bright in the NIRCam LW, it is barely visible 
in the SW. Second, the bright 870~$\mu$m source (corresponding to the arc) has 
a much fainter companion source that is 1\farcs16 away, which is likely its 
counter image. While we cannot rule out the possibility that this fainter 
companion could be due to the foreground lens, interpreting it as the counter
image is consistent with our analysis presented in the remainder of this paper.

\section{Photometry of COSBO-7}

\subsection{Photometry in the JWST NIRCam and MIRI images}

To obtain an accurate photometry of COSBO-7's counterparts in the JWST images, 
it is necessary to subtract the extended foreground lens. For this purpose, we 
utilized the {\sc Galfit} software \citep[][]{Peng2002, Peng2010} to model this 
foreground galaxy. 

\begin{figure*}
    \centering
    \includegraphics[width=\textwidth,height=\textheight,keepaspectratio]{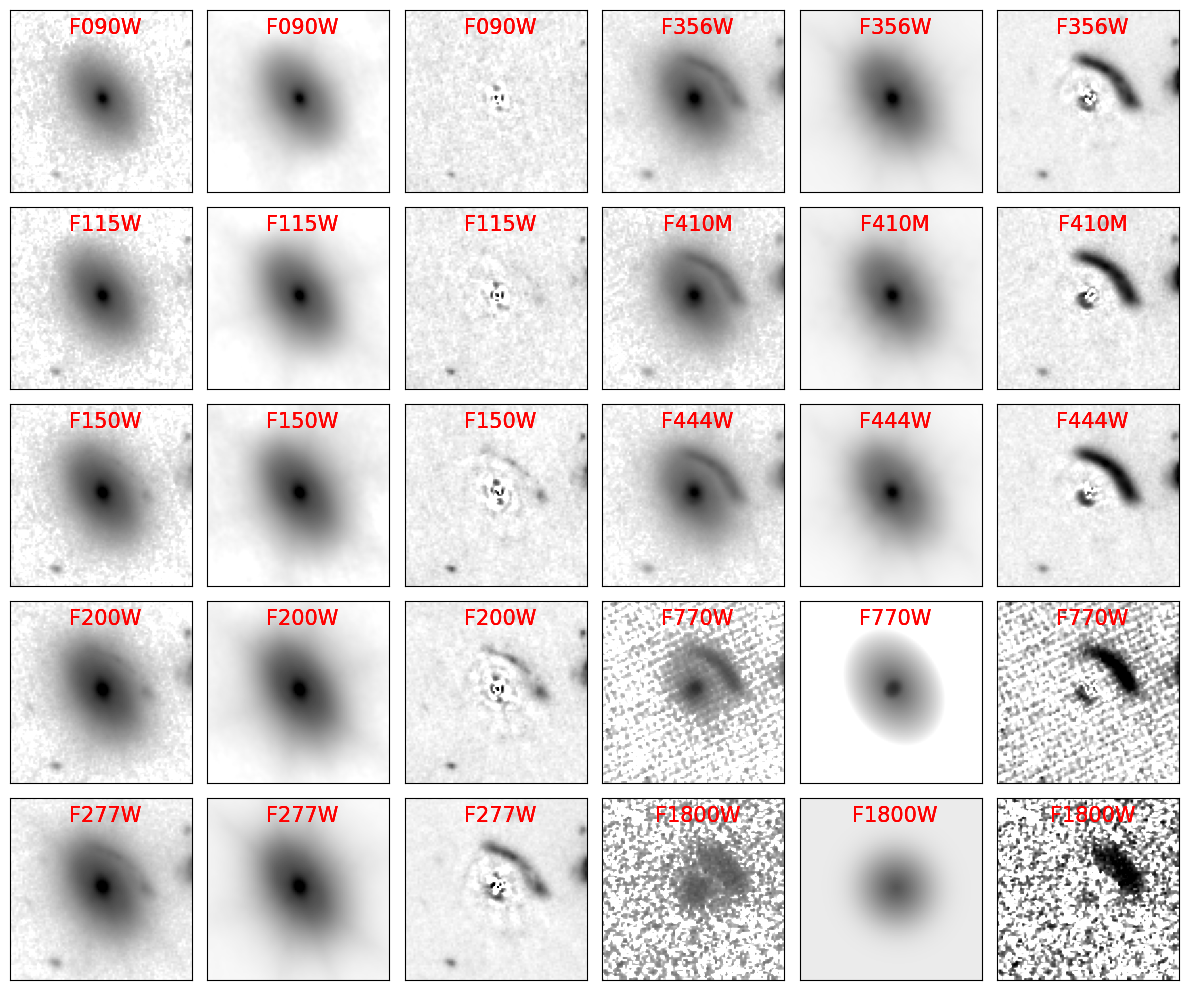}
    \caption{Removal of the foreground lens galaxy using GALFIT in the eight 
    NIRCam and two MIRI bands. For each band, the original image, the best-fit 
    model from GALFIT and the residual image after the model subtraction are
    shown from left to right. The MIRI bands only require one (F1800W) or two
    (F770W) S\'ersic profiles, while the NIRCam bands require a four-component
    model (see Section 3.3 for details).}
    \label{fig:galfit}
\end{figure*}    
        
Running {\sc Galfit} needs the point spread function (PSF) of the image to be 
analyzed. For the NIRCam images, we constructed the empirical PSFs following 
the methodology of \citet[][]{LY2022}. For a given band, we selected isolated 
stars and made cutouts of $201\times 201$ pixels centered on them. The sources 
around the stars were masked, and the cutout images were subsampled to a finer 
grid by 10 times. The centers of these cutouts were also aligned in this 
process. The fluxes of the stars were normalized to unity, and the normalized 
images were stacked using median. Finally, the stacked star image was rebinned 
by a factor of 10 in both dimensions to restore the original resolution and was 
adopted as the PSF image. For the MIRI F770W and F1800W images, we directly 
used the WebbPSF to obtain their PSFs. 

The {\sc Galfit} modeling of the foreground lens was done on 
10\arcsec~$\times$~10\arcsec\ cutouts. In the MIRI F1800W band, this galaxy was
successfully fitted by using a single S\'ersic profile
\citep[][]{Sersic1963, Sersic1968}. In the MIRI F770W band,
we had to use two S\'ersic profiles simultaneously in order to obtain a good 
fit. In the NIRCam bands, however, fitting two S\'ersic profiles would still
leave a strong residual that shows a barred spiral structure with a ring. In 
the end, we had to evoke a four-component model: a S\'ersic profile for the 
disk component, a logarithmic-hyperbolic tangent spiral (log-tanh) profile for 
the barred spiral structure, a truncated S\'ersic profile for the ring 
structure, and a point source at the galaxy's center for its compact bulge. 
This four-component model was successful in fitting the foreground galaxy in the
eight NIRCam image. Figure \ref{fig:galfit} shows the original images, the 
{\sc Galfit} models and the residual images in all the JWST bands. 

In addition to the cleaned arc, the counter image of the arc is now also 
recovered in the residual images. 
The arc is invisible in F090W, starts to appear in F115W, gets
increasingly bright through the successively redder NIRCam bands, and stays 
prominent in the two NIRCam bands up to 18~$\mu$m. Its counter image has a
similar behavior, except that it is invisible in all NIRCam SW bands, presumably
due to its much fainter nature.

The photometry was done on these residual images. For the NIRCam images, we 
carried out PSF-matching to the angular resolution of the F444W image 
(PSF full-width-at-half-maximum 0\farcs16) by convolving them with the 
convolution kernels created using 
the software {\sc pypher} from the PSF images. Matched-aperture photometry 
was then done by using {\sc SExtractor} in the dual-image mode, where the 
F444W image was set as the detection image. We took the isophotal magnitudes 
(``MAG\_ISO'') to measure the colors and corrected the normalization of the
SED to total magnitudes based 
on the difference between ``MAG\_AUTO'' and ``MAG\_ISO'' measured in the F444W 
band. Photometry in the MIRI bands was also done using {\sc SExtractor},
however no PSF-matching was done on these images because they have much coarser 
resolutions as compared to the NIRCam images. Instead, we took the 
MAG\_AUTO results in these two bands as the total magnitude measurements and 
appended them to the SED. In all cases, the errors were measured 
on the root-mean-square (rms) maps. 
The rms map of a given band was calculated using the ``WHT'' extension
of the pipeline product in this band following $rms=s/\sqrt{WHT}$,
where $s$ is the scaling factor that takes into account the 
artificial suppression of noise in the science image due to pixel resampling.
To derive this factor, we used the \texttt{astroRMS} software tool
\footnote{Courtesy of M. Mechtley; see \url{https://github.com/mmechtley/astroRMS}}, 
which is based on the algorithm that calculates the auto-correlation of the 
science image pixels (M. Dickinson, priv. comm.).

As the removal of the foreground lens the
subtraction of a smooth, noiseless model from GALFIT, this process does not 
increase the rms at the source location.
The final JWST photometric results are listed in 
Table \ref{tab:jwstphot}. 
The arc is invisible in F090W, and no flux was extracted in this band.
In other words, it is a potential F090W dropout. Following the practices in 
dropout search, a 2~$\sigma$ limit was placed in this band. To investigate
the impact of this limit in the follow-up analysis, we adopted two possible 
choices. One was the average 2~$\sigma$ depth of the image over the entire 
field as measured within an $r=$~0\farcs2 aperture, which is  28.43~mag
\footnote{The local 2~$\sigma$ depth within a $r=$~0\farcs2 aperture at the
source location is 28.72~mag, which is deeper because this region is in a
overlapped area between two tiles.}. 
Given its knotty morphology in F115W, if it were emitting in the bluer F090W
band, the emission would likely be confined within a small knot that could be
encompassed by such an aperture. Another choice was the 2~$\sigma$ limit
measured within the same MAG\_ISO aperture as used in other bands, which is 
 26.98~mag and likely very conservative.

   We did not attempt to obtain accurate photometry of
the counter image because it is much fainter and is severely affected by the
noise due to the image subtraction. Nevertheless, a crude photometry shows that
the flux ratio between the arc and its counter image is $\sim$11.7 in 
F444W.

   Following similar procedures, we also obtained the photometry 
for the foreground galaxy using its model images created by {\sc Galfit}. The
results are also reported in Table~\ref{tab:jwstphot}.

\begin{deluxetable}{ccc}
    \tablewidth{0.45\textwidth} 
    \tablecaption{JWST Photometry of the Arc and the Lens\label{tab:jwstphot}}
    \tablecolumns{3}
    \tablehead
            {
        		\colhead{Band} & \colhead{$\rm mag_{arc}$} & \colhead{$\rm mag_{lens}$}
            }
    
            \startdata	
                     F090W & $\le 28.43/26.98$      & 19.927 $\pm$ 0.003 \\
                     F115W & 25.28 $\pm$ 0.10  & 19.451 $\pm$ 0.002\\
                     F150W & 24.75 $\pm$ 0.05  & 19.076 $\pm$ 0.001\\
                     F200W & 23.78 $\pm$ 0.02  & 18.835 $\pm$ 0.001\\
                     F277W & 23.11 $\pm$ 0.01  & 18.922	$\pm$ 0.001\\
                     F356W & 22.38 $\pm$ 0.01 & 	19.438 $\pm$ 0.001\\
                     F410M & 22.08 $\pm$ 0.01 & 19.572	$\pm$ 0.001\\
                     F444W & 21.90 $\pm$ 0.01 & 	19.61 $\pm$ 0.001\\
                     F770W & 21.59 $\pm$ 0.01 & 20.916	$\pm$ 0.003\\
                     F1800W & 21.43 $\pm$ 0.04 & 21.372 $\pm$  0.015\\
            \enddata
        
    \label{table:jwstphot}
    \tablenotetext{}{Notes. The magnitudes quoted for the arc are the measured 
    values without the lensing magnification correction. Two upper limits
    in F090W are quoted, one being the averaged 2~$\sigma$ depth of the image
    measured within an $r=$~0\farcs2 circular aperture and the other being
    the 2~$\sigma$ limit measured within the MAG\_ISO aperture defined in F444W.}
\end{deluxetable}

\subsection{Photometry in the ALMA images}

\begin{figure}
    \centering
    \includegraphics[width=0.45\textwidth,height=\textheight,keepaspectratio]{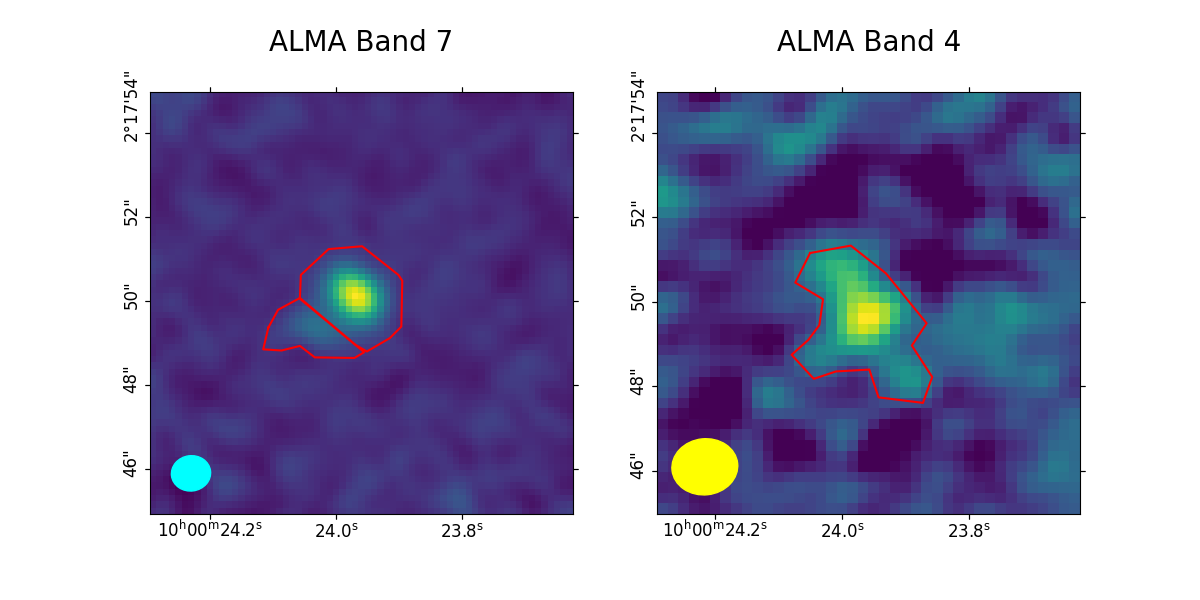}
    \caption{Polygon apertures used for photometry in the ALMA images, indicated
    in red. In Band 7, the main image and its counter image are separated. In
    Band 4, the two are not separable, and a larger polygon aperture is adopted.
 }
        \label{fig:almaphot}
\end{figure}

To maximize the signal-to-noise ratio of the extraction, the photometry in
the ALMA images was done using {\sc CASA} with polygon apertures as shown in 
Figure \ref{fig:almaphot}. Basically, this was to sum up the pixels within the
polygons. In Band 7, we separated the main image and its counter image 
(see Section 2.4) as indicated in the figure, and 
got $8.87\pm 0.89$ and $1.68\pm 0.17$~mJy, respectively. In Band 4, however, the 
counter image is completely blended
with the main image and not separable; therefore we used a large polygon for 
photometry, which gave $0.62\pm 0.07$~mJy
\footnote{The peak flux density in the Band 4 image is 0.45 mJy, which is
significantly lower than that measured in the polygon. This supports that the
Band 4 image is resolved.}.
The flux error was estimated to 
include the rms of the pixel values within the aperture and 
the uncertainty of the flux calibration 
\citep[10\% of the flux,][]{2014Msngr.155...19F}, which
were added in quadrature. As it turned out, the latter dominates the error.

\begin{deluxetable*}{cccc}
    \tablecaption{ALMA and Herschel Photometry}
    \tablewidth{0.4\textwidth}
    \tablehead{Source & R.A. & Decl. & Flux density (mJy)}
        \startdata
        ALMA 870~$\mu$m & 10:00:23.971 & 02:17:50.097 & 8.87 $\pm$ 0.89 \\ 
                        & 10:00:24.032 & 02:17:49.373 & 1.68 $\pm$ 0.17 \\ 
        ALMA 2.07~mm    & 10:00:23.962 & 02:17:49.598 & 0.62 $\pm$ 0.07 \\ 
        SPIRE 250~$\mu$m & ...         & ...          & 15.58 $\pm$ 5.36 \\
        SPIRE 350~$\mu$m & ...         & ...          & 26.03 $\pm$ 9.68 \\
        SPIRE 500~$\mu$m & ...         & ...          & 19.24 $\pm$ 5.05 
        \enddata   
        \label{tab:almaherschel}
    \tablenotetext{}{Notes. 
    The main image and its counter image are separable in
    ALMA 870~$\mu$m, and the corresponding results are given in the first and 
    the second row, respectively. Their positions are those of the peaks.
    The position in the Herschel SPIRE images is fixed to the one measured in 
    the VLA 3~GHz (see Table \ref{tab:literature}), which is consistent with 
    the ALMA position of the main image.}
\end{deluxetable*}

\subsection{Photometry in the Herschel SPIRE maps}

\begin{figure}
    \centering
    \includegraphics[width=0.5\textwidth,height=\textheight,keepaspectratio]{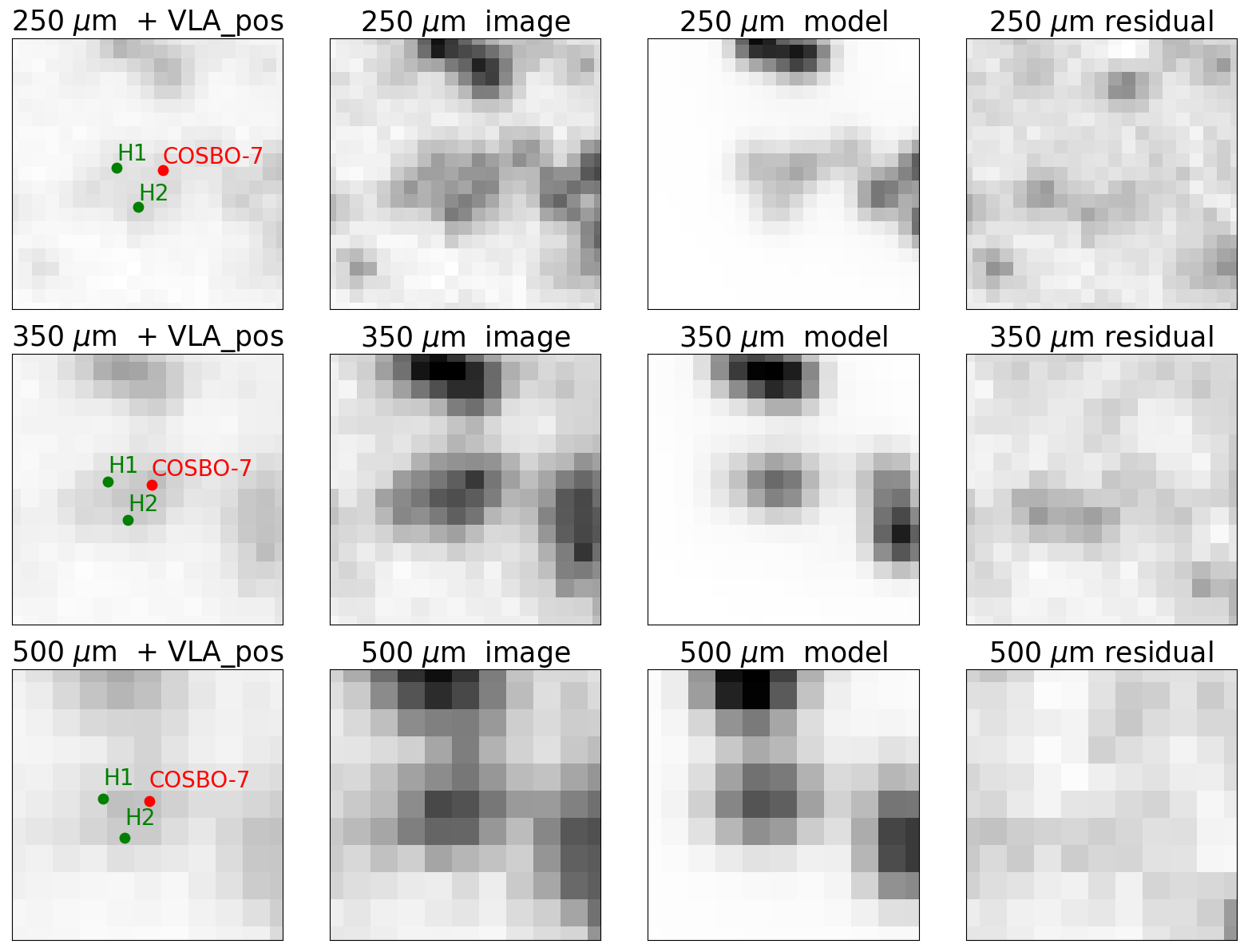}
    \caption{De-blending of the Herschel SPIRE images. From
    top to bottom, the three rows show the case in 250, 350 and 500~$\mu$m, 
    respectively. The images are 2\arcmin\ on a side.
    The first panel shows the VLA positions of
    the three blended sources overlaid on the SPIRE image: COSBO-7 (red symbols)
    H1 and H2 (green symbols). The second to fourth panels show the original
    image, the GALFIT-constructed model image by fitting PSFs at the fixed
    VLA positions, and the residual image after subtracting the model from
    the original image. }
    \label{fig:spiregalfit}
\end{figure}

   Due to their coarse spatial resolutions, the SPIRE images of COSBO-7 are 
severely blended with its neighbors, which can be seen in Figure 
\ref{fig:spiregalfit}. In the 250~$\mu$m image, COSBO-7 is blended with at least
two neighbors (H1 and H2) that can also be identified with the VLA sources. H2
is indistinguishable in 350~$\mu$m and seems disappeared from 500~$\mu$m,
however H1 could persist in all three bands. To de-blend, we fit the SPIRE 
PSFs 
\footnote{The SPIRES PSFs of 1\arcsec\ pixel scale were retrieved from \url{http://archives.esac.esa.int/hsa/legacy/ADP/PSF/SPIRE/SPIRE-P/}. 
They were then subsampled/rebinned to 5\farcs00, 8\farcs33, and 10\farcs00\ to 
be used by GALFIT, which match the pixel scales of the HerMES 250, 350 and 
500~$\mu$m maps, respectively.} 
to the maps using {\sc Galfit} at the VLA positions of COSBO-7, H1 and H2 
simultaneously. As it turned out, de-blending all three sources were successful
in 250 and 350~$\mu$m, with H2 of negligible flux in the latter band as 
expected. However, {\sc Galfit} crashed if fitting all three simultaneously in
500~$\mu$m and would only run without fitting H2. Therefore, we discarded H2 
when de-blending in this band. To convert from Jy~beam$^{-1}$ (the unit of the
SPIRE maps) to mJy, we used the beam solid angles of 469.35, 831.27 and
1804.31~arcsec$^2$ for 250, 350 and 500~$\mu$m, respectively, which are the
values adopted by the Herschel SPIRE data reduction pipeline to produce these
maps. To summarize, COSBO-7 has flux density of $16.58\pm5.36$, $26.03\pm9.68$
and $19.24\pm5.05$~mJy in 250, 350 and 500~$\mu$m, respectively. The quoted
errors include the confusion noises of 3.32, 6.51 and 4.91~mJy (based on the
HerMES final data release) in these three bands, respectively.

\section{SED Fitting}

\begin{figure*}
    \centering
    \includegraphics[width=\textwidth,height=\textheight,keepaspectratio]{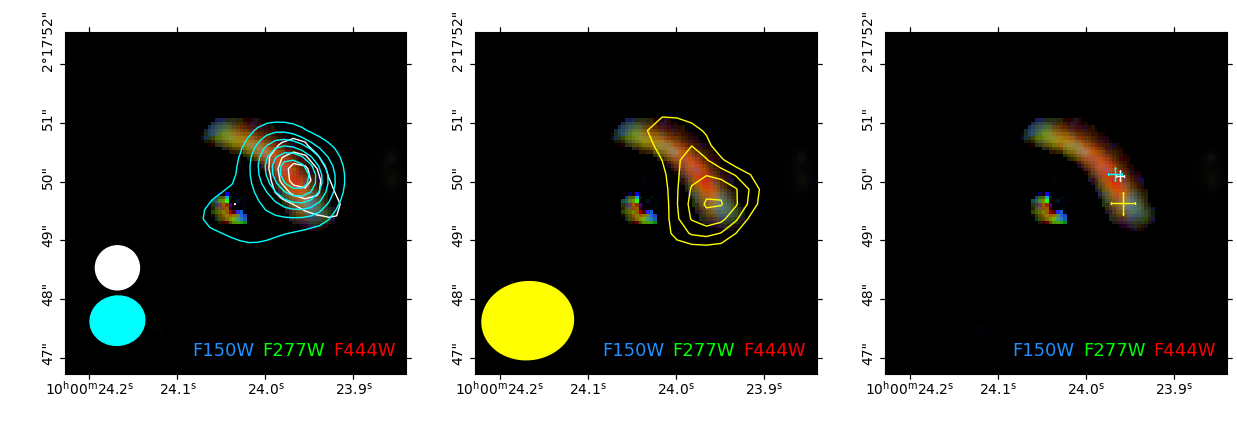}
    \caption{Comparison of the dust emission as detected by ALMA and the star
    light as detected by JWST NIRCam. The background image is a color composite
    of the NIRCam residual images in F150W (blue), F277W (green) and F444W 
    (red) after the removal of the foreground lens. The left panel shows the
    contours of the ALMA Band 7 image (cyan), which coincide the contours of
    the VLA 3~GHz image (white). The middle panel shows the contours of the ALMA
    Band 4 image (yellow). The peak positions of the ALMA Band 7 and 4 emissions
    as well as that of the VLA 3~GHz emissions are marked as the ``+'' symbols
    in the right panel using the same color coding.
 }
        \label{fig:nircam_alma}
\end{figure*}

\subsection{Distribution of dust-embedded region} 

Figure \ref{fig:nircam_alma} compares the NIRCam and the ALMA images by
superposing the ALMA 870~$\mu$m and 2.07~mm contours on the NIRCam residual 
image (i.e., after the removal of the foreground lens). 

The comparison of the NIRCam and the ALMA images indicates that the 
dust-embedded star-forming region giving rise to the far-IR-to-mm emission does
not cover the entire stellar population of the galaxy. This is shown in Figure
\ref{fig:nircam_alma}, where the ALMA 870~$\mu$m and 2.07~mm contours are
superposed on the NIRCam color-composite image obtained after the removal of 
the foreground lens. Clearly, the arc has a more extended stellar light 
distribution than the dust emission: the 870~$\mu$m emission on the arc is
concentrated on its southwestern part
\footnote{The systematic difference between the ALMA positions and the CANDELS
HST F160W positions in the COSMOS field that our astrometry is tied to is only
$\sim$0\farcs06--0\farcs07 \citep[][]{LY2022} and cannot account for the
offset seen here. In addition, the offset cannot be due to the random error of
the ALMA position, because the high signal-to-noise ratio of the ALMA emission
means that the centroid is accurate to $\sim$0\farcs06 \citep[see also][]{LY2022}.}, 
and so is the 2.07~mm emission.

There is another piece of evidence supporting a confined dust-embedded 
star-forming region in this galaxy, which comes from the flux ratio of 
the arc and its counter image. As mentioned in Section 3, this ratio is 
$11.7:1$ in the NIRCam F444W and is $5.3:1$ in the ALMA Band 7. Had the dusty
region is well mixed with the underlying stellar population throughout the
galaxy, the two ratios would be the same. Such a large discrepancy can be
explained if the dusty region is confined to a limited area within the galaxy,
which does not share the same magnification as the whole galaxy.

For this reason, we carry out SED fitting separately for the stellar population 
as detected in near-to-mid-IR and the dust-embedded starburst region as
detected in far-IR-to-mm.

\begin{deluxetable*}{ccccccccccc}
    \centering
    \tablecaption{Physical properties of COSBO-7 from SED fitting}
    \tablecolumns{11}
     \tablehead{
     Software & $z_{\rm ph}$ & $z_{\rm best}$                                & $\chi^2$ & log$\rm\frac{\mu M_*}{M_\odot}$    & log($\rm \frac{\mu SFR}{M_\odot~yr^{-1}}$)               & Age(Gyr)    & $A_{\rm V}$                  & E(B-V) & $Z_* $ & $\tau$(Gyr) 
     }
     \startdata
     LePhare  & $7.1^{+0.8}_{-0.0}$ & 7.1  & 11.3 & $11.16^{+0.08}_{-0.41}$  & $4.85^{+0.77}_{-0.56}$  & 0.1 & ...  & 0.7  & 0.02  & 0.01 \\ 
     EAZY  & $7.26^{+0.35}_{-0.11}$ & 7.23 & 13.8 & $11.54 ^{+0.03}_{-0.03}$  & $2.68^{+0.07}_{-0.06}$  & ...   & $0.69^{+0.8}_{-0.09}$  & ...  & ... & ... \\ 
     X-CIGALE & $7.54^{+0.25}_{-0.25}$  & 7.3 & 19.3  & $11.14^{+0.21}_{-0.43}$  & $3.53^{+0.24}_{-0.61}$ & 0.71 & ... & 0.5  & 0.02 & 0.14  \\ 
     Bagpipes & $7.67^{+0.03}_{-0.03}$ & 7.67 & 20.0 & $10.56^{+0.08}_{-0.05}$ & $2.56^{+0.06}_{-0.05}$ & 0.01 & $1.93^{+0.04}_{-0.05}$ & ... & 0.07 & 4.96 \\ 
     \hline
     LePhare  & $7.0^{+0.8}_{-0.2}$ & 7.0  & 11.3 & $11.51^{+0.11}_{-0.41}$  & $4.54^{+1.04}_{-0.28}$  & 0.07 & ...  & 0.7  & 0.02  & 0.01 \\ 
     EAZY  & $6.97^{+0.21}_{-0.29}$ & 7.02 & 12.0 & $11.55^{+0.08}_{-0.13}$  & $2.63^{+0.24}_{-2.18}$  & ...   & $0.77^{+0.23}_{-0.10}$  & ...  & ... & ... \\ 
     X-CIGALE & $7.03^{+1.1}_{-1.1}$  & 7.2 & 19.1  & $11.08^{+0.22}_{-0.48}$  & $3.46^{+0.25}_{-0.73}$ & 0.83 & ... & 0.5  & 0.02 & 0.15  \\ 
     Bagpipes & $6.89^{+0.08}_{-0.14}$ & 6.89 & 27.6 & $11.13^{+0.25}_{-0.25}$ & $2.98^{+0.09}_{-0.18}$ & 0.16 & $1.90^{+0.07}_{-0.13}$ & ... & 0.05  & 5.25 \\ 
   \enddata
   \label{tab:sedfit}
   \tablenotetext{}{Notes. 
   The top four rows are for the SED using the F090W limit of 28.43~mag, while the bottom four rows are for that using the F090W limit of 26.98~mag.
   For LePhare, EAZY (EAZY+FAST) and X-CIGALE, the $z_{\rm ph}$ 
   and $z_{\rm best}$ values are the mean photometric redshift weighted by 
   $P(z)$ and that of the best-fit template, respectively.  For Bagpipes, 
   $z_{\rm ph}$ is the 50th percentile value with errors indicating the 16th 
   and 84th percentiles, and $z_{\rm best}$ is the average of the 16th to 84th 
   percentile. $\chi^2$ is the total (not reduced) value corresponding
   to $z_{\rm best}$. The estimates of stellar mass ($M_*$) and SFR are affected
   by the magnification factor $\mu$, and the quoted values 
   ($\mu M_*$ and $\mu$SFR) are not de-magnified. The age derived by Bagpipes
   is the mass-weighted age. The dust extinction values are
   either given in terms of $A_{\rm V}$ or $E(B-V)$, depending on the software 
   tool in use; for the Calzetti's extinction law, 
   $A_{\rm V}\approx 4.04\times E(B-V)$.
   $Z_*$ is metallicity, which is fixed to the solar metallicity in the LePhare 
   run but is a free parameter in the X-CIGALE and Bagpipes runs. 
   $\tau$ is the characteristic time scale of the
   exponentially declining SFH.
   }
\end{deluxetable*}

\subsection{Fitting of the SED based on JWST photometry}

   \citet[][]{Pearson2024} fit to the combined SED that is the mixture of 
the emissions from both the arc and the foreground lens and obtained 
$z_{\rm ph}=3.4\pm0.4$ for the arc. Their SED incorporates the data 
from the Sloan Digital Sky Survey to MAMBO; in terms of the JWST data, however,
they only used the two-band MIRI imaging data. Obviously, there is a vast room 
for improvement in both the data utilization and the fitting method.
We performed a full analysis of the SED of the arc, using the JWST photometry
summarized in Table~\ref{tab:jwstphot}. Considering that the removal of the 
foreground galaxy is not perfect, we added (in quadrature) 10\% of the fluxes 
to the errors.

   It is well known that SED fitting results depend on the fitting software as
well as the models that it employs, and therefore we utilized four different 
tools, namely, LePhare \citep[][]{Arnouts1999, Ilbert2006}, 
the python version of EAZY \citep[EAZY-py;][]{Brammer2008},
Bagpipes \citep[][]{Carnall2018}, and 
CIGALE \citep[][]{Burgarella2005, Noll2009, Boquien2019}. The redshift was 
allowed to vary from 0 to 10, and no prior was applied. 

  The fitting by LePhare used the population synthesis models of 
\citet[][BC03]{Bruzual2003} with solar metallicity and the initial mass 
function (IMF) of \citet[][]{Chabrier2003}. The templates were constructed 
assuming exponentially declining star formation histories (SFHs) in the form of 
SFR~$\propto e^{-t/\tau}$, where $\tau$ ranged from 0 to 13 Gyr (0 for simple
stellar population and 13 Gyr to approximate constant star formation).
LePhare allows the use of magnitude limits, which we enabled so that
any solutions violating the F090W limit were rejected.
In the EAZY run, we used the ``tweak\_fsps\_QSF\_12\_v3'' templates, which 
are a modified version of the original Flexible Stellar Population Synthesis 
models \citep[FSPS;][]{Conroy2010} tailored for galaxies at high redshifts
\citep[][]{Finkelstein2022, Larson2023}. These 
templates use the Kroupa IMF \citep[][]{Kroupa2001}. 
For CIGALE, we used the 2022.1 version \citep[``X-CIGALE'';][]{YangCIGALE2022}
of the program and the BC03 model with the Chabrier IMF.
The adopted SFHs were the delayed-$\tau$ model, with other settings similar to 
the SFHs used in the LePhare run. The age of the stellar population was limited
to $>10$~Myr, and the AGN contribution was set to 0. The metallicity was
allowed to vary between  $0 \le Z_{*}/Z_{\odot} \le 2.5$. The range of the
ionization parameter was set to $-3\le {\rm log}(U)\le -2$. 
For the Bagpipes run, we also used the delayed-$\tau$ model and other
settings similar to the CIGALE run, with the exception that the ionization
parameter was fixed to ${\rm log}(U)=-3$.

We note that its 
underlying stellar population synthesis models are those of BC03 with the 
Kroupa IMF. For LePhare, CIGALE and Bagpipes, we adopted the Calzetti's dust 
extinction law \citep[][]{Calzetti1994, Calzetti2001} with $E(B-V)$ ranging 
from 0 to 1.0~mag. The templates used in the EAZY run have also adopted the 
same extinction law but the amount is hardwired within.

EAZY, CIGALE and Bagpipes do not directly use magnitude limits; following a
common practice, we set the flux density and its error in F090W to the 
magnitude limit quoted in Table~\ref{tab:jwstphot}.

\begin{figure*}
    \centering
    \includegraphics[width=0.9\textwidth,height=\textheight,keepaspectratio]{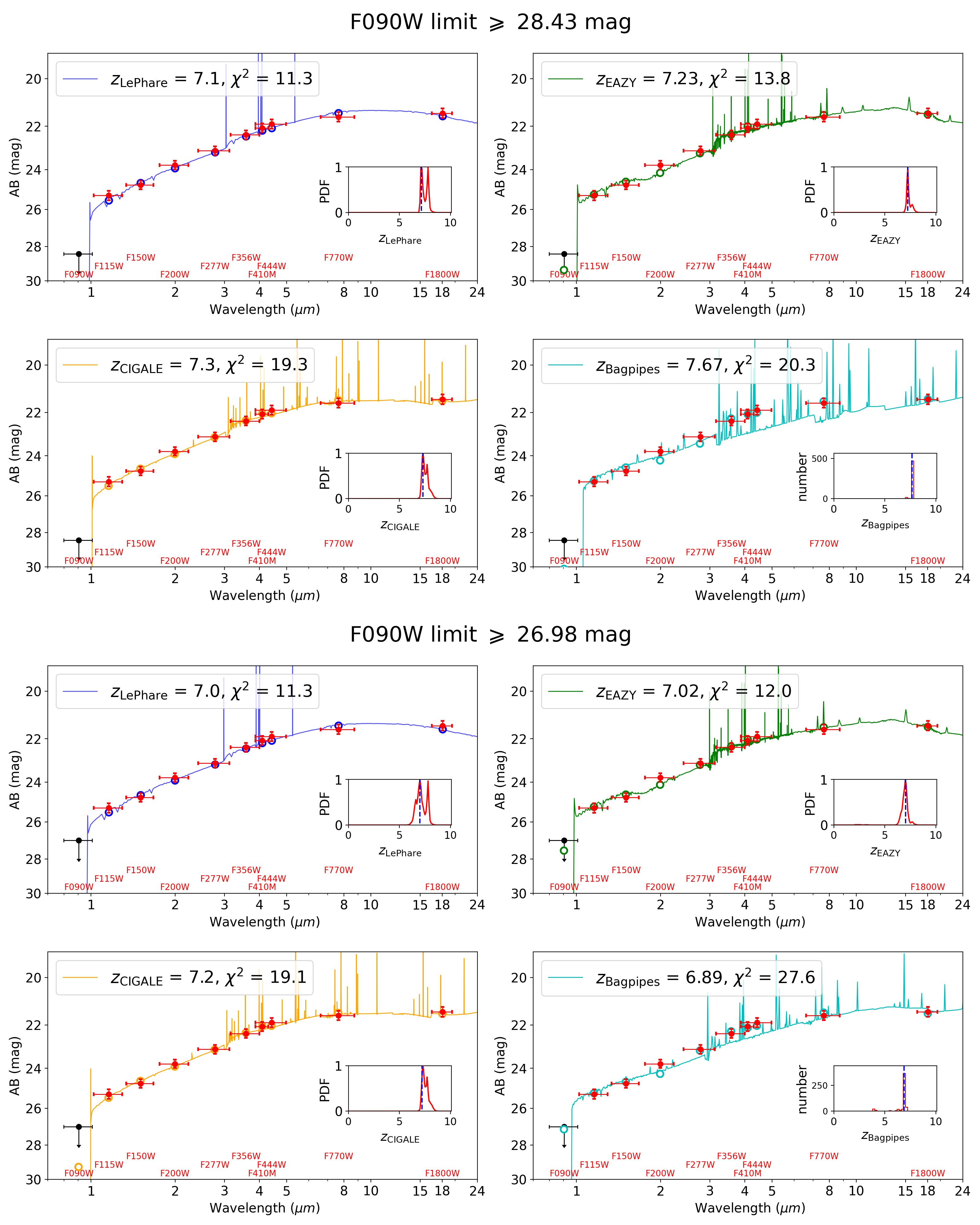}
    \caption{Fitting of COSBO-7's optical-to-mid-IR SED to derived its 
    photometric redshift based on four different SED fitting tools. The data
    points (solid symbols) and upper limit are from the JWST photometry listed
    in Table~\ref{tab:jwstphot}, and the corresponding passbands are labeled at
    the bottom. 
    The cases for the F090W upper limit of 28.43~mag and 26.98~mag 
    are shown in the top four and the bottom four panels, respectively.
    In the cases of LePhare, EAZY and X-CIGALE,
    the superposed spectrum is that of the best-fit template, and the inset
    shows the probability distribution function (PDF) of redshift ($P(z)$).
    In the case of Bagpipes, the spectrum is the 50th percentile posterior  
    spectrum and the inset shows the distribution of the posterior redshifts.
    In all cases, the open symbols are the synthesized magnitudes of the
    superposed spectrum at the corresponding bands. See Section 4.2 for details.
    }
    \label{fig:sedjwst}

\end{figure*}

   The results from these four different tools are given in 
Table~\ref{tab:sedfit} separately for the two choices of the F090W upper limit.
Interestingly, they suggest that a high redshift of COSBO-7 is very
likely, which ranges from $\sim$6.9 to 7.7. This is demonstrated in 
Figure~\ref{fig:sedjwst} using the results corresponding to the 
``$z_{\rm best}$'' column in Table \ref{tab:sedfit}.
Of course, we cannot rule out the low-redshift possibility due to the nature of
SED fitting. 
In Appendix~\ref{forcelowz}, we present the fitting results when forcing
the redshift to $z\leq 6$. These solutions, however, all have significantly 
worse $\chi^2$, with $\Delta \chi^2 \approx 16$--65 and 6--9 when
adopting 28.43 and 26.98~mag as the F090W upper limit, respectively.

Therefore, it is fair to say that our results are more 
in favor of a high-redshift interpretation at $z > 7$.
During the revision of this paper, a new set of ALMA Band 3 spectroscopic
data became publicly available. We detected a single line in these data, which 
could be the CO (7-6) line at $z=7.455$ (see Appendix \ref{almab3}).
Such a redshift, if confirmed, will put COSBO-7 among
the earliest DSFGs in the universe known to date. 

    Another interesting point is that the derived stellar mass is consistent
among the four sets (within $\sim$1 dex) but the SFR varies by 1--2~dex. The 
discrepant SFR derivations are largely due to the different templates in use. 
Firstly, the inclusion of nebular continuum emission matters. The LePhare run 
is the only one whose templates do not include nebular continuum emission
(albeit with {\it ad hoc} line emission), and 
therefore the UV light must be accounted for by using only stars, which tends to
result in a high SFR estimate. Secondly, and probably more importantly, the 
adopted SFH plays a significant role. This is because explaining the strong UV 
emission of our object must involve a large number of high-mass stars of very
young ages, and the adopted SFHs can result in large differences when confined 
to a short time interval due to the young age of the universe at the derived
redshift. The SFH of the BC03 templates used in the LePhare run 
has an exponentially declining SFR (the classic ``$\tau$'' model). The small 
$\tau$ (10~Myr) and young age (70--100~Myr) of the best-fit template means that
the inferred stellar population is very close to an instantaneous burst at its 
beginning, and therefore it must have a very high SFR. Indeed, the LePhare run 
gave the highest SFR estimate. 
The X-CIGALE run, which had the same underlying
BC03 model as the LePhare run, gave 0.9--1.3~dex lower SFR (the second 
highest). 
This can be attributed to its delayed-$\tau$ SFH, which stretches the star 
formation process to a longer time scale (i.e., not so close to an 
instantaneous burst), leading to a smaller SFR. 
Similar arguments can be made to explain the SFR derived by the Bagpipes 
run.

The EAZY-py run gave a comparable SFR estimate as the 
Bagpipes run, and this is mostly because the adopted SFH spreads the 
formation of stars to a few discrete events. 

    For the sake of completeness, we also fit the SED of the lens. The results
are given in Appendix \ref{sedlens}.

\subsection{Fitting of the far-IR-to-mm SED}

   We used a modified black body (MBB) model to fit the far-IR-to-mm SED
constructed from the Herschel and ALMA photometry in 
Table~\ref{tab:almaherschel} and the COSBO MAMBO-2 measurement at 1.2~mm in 
Table \ref{tab:literature}. Other measurements in Table \ref{tab:literature} 
are not included in the SED for various reasons: the AzTEC 1.1~mm result is of 
low S/N, the SCUBA-2 450~$\mu$m result is likely contaminated by an unrelated 
close neighbor (as judged from the Herschel SPIRE images) and cannot be 
de-blended, and the SCUBA-2 850~$\mu$m results are superseded by the ALMA 
870~$ \mu$m measurement. In this regime, the ALMA 870~$\mu$m image is the only 
one of a high enough resolution that separates the two lensed images of 
COSBO-7, and  we combined the two so that the result in this band could be used 
together with those at other wavelengths. The combined flux density is 
$S_{870}=10.55\pm1.06$~mJy.

   Following \citet[][]{MY2015} and \citet[][]{Yan2016}, we used the 
single-temperature MBB model in this form:
\begin{equation}\label{eq:mbb}
\begin{split} 
     S_{\lambda}(\lambda) & \equiv N \cdot I_{\rm mbb}(\lambda)\\
                & = N\frac{1-\mathrm{e}^
         {-(\frac{\lambda_0}{\lambda})^{\beta}}}{1-\mathrm{e}^{-1}}
     \frac{(\frac{2hc^2}{\lambda^5})}{\mathrm{e}^{hc/(\lambda kT_{\rm mbb})}-1} \,,
\end{split}
\end{equation}
where $T_{\rm mbb}$ is the characteristic temperature of the MBB, $N$ is the
scaling factor that is related to the luminosity, $\beta$ is the
emissivity, and $\lambda_0$ is the reference wavelength where the opacity is
unity. We set $\lambda_0=100$~$\mu$m. As $\beta$ affects $T_{\rm mbb}$, 
the fit was done using three different choices of $\beta=1.5$, 2.0 and 2.5.
We fixed the redshift to $z=7.455$ (see Appendix A). 

   The results are shown in Figure \ref{fig:mbbfit}. Regardless of the choice 
of $\beta$, the fit gives a high dust temperature of $T_{\rm mbb}>90$~K.
This is significantly higher than the dust temperature of $\sim$40--60~K often 
seen in the SMGs at $z\approx 2$--4. 

   The reported $L^{\prime}_{\rm FIR}$ values in Figure \ref{fig:mbbfit} are 
the far-IR luminosity integrated from 60 to 1000~$\mu$m in the rest-frame. Note 
that the fit was done by combining the two lensed images. Assuming that the 
split of $L^{\prime}_{\rm FIR}$ between the two follows the flux density ratio 
of 5.3 in ALMA 870~$\mu$m (see Table~\ref{tab:almaherschel}), the main image 
(corresponding to the arc seen in the JWST images) has 
$\mu_d L_{\rm FIR}=6.7\times 10^{12} L_\odot$ for $\beta=1.5$ and 
$8.4\times 10^{12} L_\odot$ for $\beta=2.0$ and 2.5. To obtain
the intrinsic $L_{\rm FIR}$, these values should be divided by the lensing 
amplification factor $\mu_d$ of the dusty region (see Section 5).

    From $L_{\rm FIR}$, one can infer the SFR of the dust-embedded 
population. Using the standard $L_{\rm IR}$ to SFR conversion of 
\citet[][]{Kennicutt1998}, i.e., 
${\rm SFR_{IR}} = 1.0\times 10^{-10} L_{\rm IR}/L_\odot$ for a Chabrier IMF,
and ignoring the difference between $L_{\rm FIR}$ and $L_{\rm IR}$, the 
$\mu_d L_{\rm FIR}$ values quoted above correspond to
$\mu_d {\rm SFR_{IR}} = 670$ and 840~$M_\odot$~yr$^{-1}$, respectively.

   Lastly, the dust mass $M_d$ can also be derived from the MBB fit, for which
we followed the recipe of the {\sc cmcirsed} code by \citet[][]{Casey2012}.
Again, the reported $M^{\prime}_d$ values in Figure \ref{fig:mbbfit} are for
the two images combined and are not corrected for the lensing magnification. 
Using the same split of $5.3:1$ as above for the two images, we obtain the dust
mass based on the main image 
$\mu_d M_d = 1.1$, 1.6, and $2.3\times 10^8 M_\odot$ for $\beta=1.5$, 2.0 and
2.5, respectively.

\begin{figure*}
    \centering
    \includegraphics[width=\textwidth,height=\textheight,keepaspectratio]{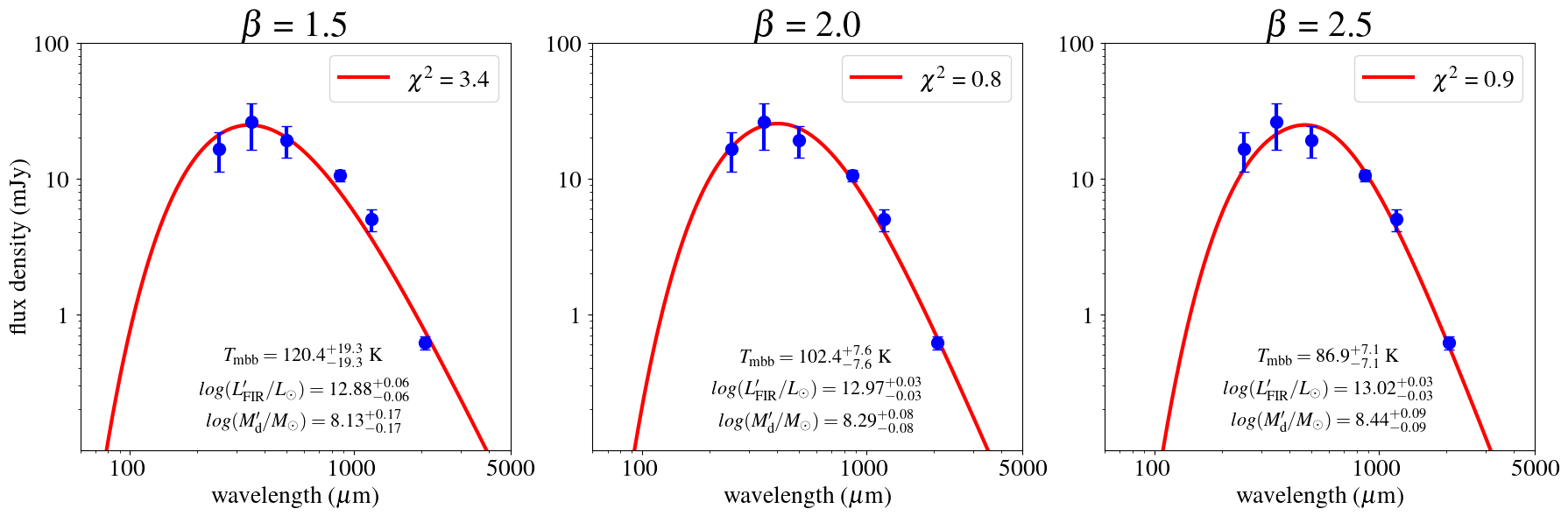}
    \caption{Single-temperature modified black body (MBB) fit of the 
    far-IR-to-mm SED using three different choices of the emissivity value
    ($\beta$) as noted. 
    The fit is done at the fixed redshift of 7.455 
    (see Appendix \ref{almab3}).
    The data points in each panel show the measurements in 
    Herschel SPIRE 250, 350 and 500~$\mu$m, ALMA 870~$\mu$m, MOMBO-2 1.2~mm, 
    and  ALMA 2.07~mm. These measurements are the combined values of the two 
    lensed images, whose flux density ratio in ALMA 870~$\mu$m is $5.3:1$. The 
    red curve is the best-fit MBB model, and the corresponding dust temperature
    ($T_{\rm mbb}$) as well as the total far-IR luminosity ($L_{\rm FIR}$) are 
    labeled. The $L_{\rm FIR}$ value is for the two lensed images combined and 
    is not de-magnified.}
    \label{fig:mbbfit}
\end{figure*}

\section{Lens modeling}

    To obtain the intrinsic luminosity and morphology of COSBO-7, we 
re-constructed its image in the source plane by modeling the residual F444W 
image where the foreground lens is subtracted. We utilized the Pipeline for 
Images of Cosmological Strong lensing \citep[PISC;][]{Li2016PICS} and the 
Affine Invariant Markov chain Monte Carlo Ensemble sampler 
\citep[Emcee;][]{DFM2013EMCEE}. The mass model of the lens was a Singular 
Isothermal Ellipsoid \citep[SIEs;][]{Kormann1994}, which is analytically 
tractable and has also been shown to be applicable in similar strong lensing 
cases
\citep[e.g.,][]{Koopmans2006, Gavazzi2007, Dye2008}. The convergence map is 
given by
\begin{equation}
    \kappa(x, y)= \frac{\theta_{\rm E}}{2}\frac{1}{\sqrt{x^2/q+y^2 q}} \;,
\end{equation}
where $q$ is the axis ratio of the lens and $\theta_{\rm E}$ is the Einstein 
radius. The environmental lensing effects are modeled with external shear 
$(\gamma^{ext}_1, \gamma^{ext}_2)$. The light distribution of the source is an 
elliptical S\'ersic profile \citep[][]{Sersic1968} following
\begin{equation}
I(R) =  I_{\rm eff}~{\rm exp} \left\lbrace -b_{n} \left[ \left( \frac{R}{R_{\rm eff}}\right)^{1/n} - 1 \right ] \right\rbrace \, ,
\end{equation}
where $R = \sqrt[]{x^2 /q+y^2 q }$, $R_{\rm eff}$ is the effective radius in 
arc-second, $I_{\rm eff}$ is the intensity at the effective radius, and 
$n$ is the index of the S\'ersic profile. Assuming that the center of mass 
aligns with the center of light and that the scaling factor $I_{\rm eff}$ can
be reduced by normalizing both the observation and source model, the strong 
lensing system can be fully described by using 
$\{\theta_{\rm E}, e^l_1, e^l_2, \gamma^{ext}_1, \gamma^{ext}_2\}$ plus 
$\{y^{s}_{1}, y^{s}_{2}, R^s_{\rm eff}, e^s_1, e^s_2, n^s\}$, where 
$\{e^l_1, e^l_2\}$ and $\{e^s_1, e^s_2\}$ are the complex ellipticity of the 
lens and the source, respectively, and $\{y^{s}_{1}, y^{s}_{2}\}$ is the 
angular position of the source in the source plane. The relation between 
$\{q, \phi\}$ and $\{e_1, e_2\}$ is given by
\begin{eqnarray}
    q & = & (1 - e)/(1 + e) \\
    \phi & = & 1/2\,{\rm arctan}(e_2/e_1)\, ,
\end{eqnarray}
where $e = \sqrt{e_1^2 + e_2^2}$. 

To fit the data, we designed the likelihood below: 
\begin{equation}
    \mathcal{L} = \sum \frac{(img^{mdl}_{i}-img^{obs}_{i})^2}{wht_{i}}msk_{i}\, ,
\end{equation} 
where $img^{obs}$ and $img^{mdl}$ are the observed image and that generated by 
the model, respectively, $wht$ is the weight map
reflecting the relative noise properties of $img^{obs}$, $msk$ is the
mask file that retains only valid pixels for the modeling, and $i$ is the 
index of pixels. We first found the best-fit results using the \texttt{optimize.minimize} function in 
\textsc{Scipy} \citep[][]{scipy2020}\footnote{\url{https://scipy.org/}}
and then used \textsc{Emcee}\footnote{\url{https://emcee.readthedocs.io/}} 
to explore the posterior distributions of the parameters of 
$\{\theta_{\rm E}, e^l_1, e^l_2, \gamma^{ext}_1, \gamma^{ext}_2, y^{s}_{1}, y^{s}_{2}, R^s_{\rm eff}, e^s_1, e^s_2, n^s, F_{scale}\}$. 
The above modeling procedure was implemented on angular scales without
assuming redshifts of the lens and the source.
Table \ref{tab:bestfits} lists the medians and 1~$\sigma$ confidential 
intervals of the posteriors. 

\begin{deluxetable*}{ccccccccccc}
    \centering
    \tablecaption{The medians and 1-$\sigma$ uncertainties of the posteriors of the lens model}
    \tablecolumns{11}
     \tablehead{
     $\theta_{\rm E}\,(^{\prime\prime})$ &
     $e^{l}_1$ & 
     $e^{l}_2$ & 
     $\gamma^{ext}_1$ & 
     $\gamma^{ext}_2$ &
     $y^{s}_1\,(^{\prime\prime})$ &
     $y^{s}_2\,(^{\prime\prime})$ &
     $r^{s}_{eff}\,(^{\prime\prime})$ &
     $e^{s}_1$ &
     $e^{s}_2$ & 
     $n^s$}
     \startdata
    $ 0.59^{+0.06}_{-0.08}$ & 
    $ 0.17^{+0.05}_{-0.07}$ &
    $ 0.14^{+0.06}_{-0.06}$ & 
    $-0.13^{+0.06}_{-0.08}$ & 
    $-0.02^{+0.10}_{-0.08}$ & 
    $ 0.44^{+0.06}_{-0.04}$ & 
    $ 0.25^{+0.03}_{-0.06}$ & 
    $ 0.16^{+0.01}_{-0.01}$ & 
    $ 0.56^{+0.11}_{-0.11}$ & 
    $-0.45^{+0.12}_{-0.12}$ & 
    $ 0.24^{+0.08}_{-0.06}$ 
    \enddata
   \label{tab:bestfits}
   \tablenotetext{}{Notes. As the modeling procedure is in angular scale, the units of $\theta_{\rm E}$, $y^s_1$, $y^s_2$, and $r^s_{eff}$ are arcsec. The estimate of each parameter is the median of the corresponding posterior, and the 1~$\sigma$ uncertainties are estimated by using the confidential interval of $[16\%,\,84\%]$.}
\end{deluxetable*}

\begin{figure*}
        \centering
        \includegraphics[width=\textwidth,height=\textheight,keepaspectratio]{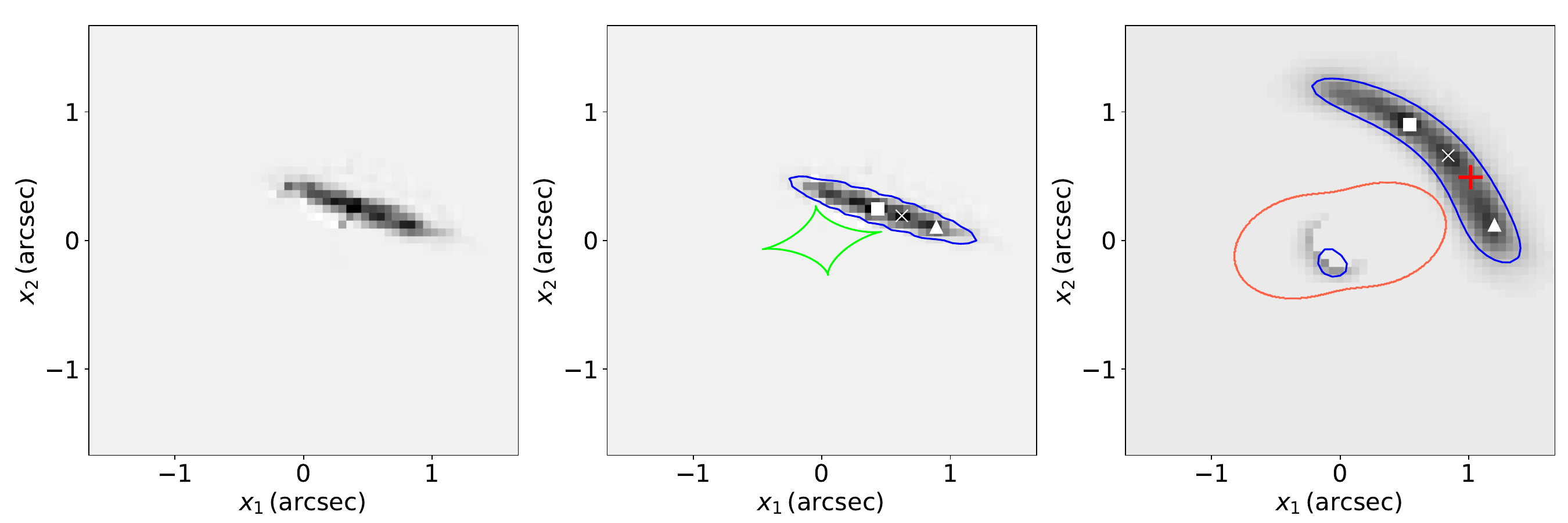}
        \caption{Lensing modeling of COSBO-7. The left panel shows the
        reconstructed image in the source plane. The middle panel overlays the 
        source model, shown as the blue contour, on the same reconstructed 
        image. The green curve represents the caustics of the lensing system. 
        The right panel superposes the modeled lensed images in blue contours on 
        the observed F444W image, and the red curve represents the critical 
        curve of the strong lensing system. The blue contours in both the middle 
        and right panels correspond to $25\%$ maximum of the F444W observation.  
        In addition, the white symbols (square, cross, and triangle) mark the 
        positions of the three clumps in the F444W data (right panel) and the 
        reconstructed source image (middle panel), respectively. The small blue 
        contour within the critical curve in the right panel is the counter 
        image of the arc predicted by our lensing model, and the red plus
        symbol indicates the peak of the main image in the ALMA 870~$\mu$m data. 
        Note that the effect of PSF smearing is included in the right panel but 
        not in the other two panels.}
        \label{fig:lensmodel}
\end{figure*}

    According to the parameters given in Table~\ref{tab:bestfits}, we generate 
the model-predicted images of the source and the lensed arc, which are shown
in Figure \ref{fig:lensmodel}. Interestingly, the best-fitted S\'ersic model
of the source has $n^s=0.24$ and an axis ratio of $\sim$0.16, suggesting that
it is a nearly edge-on disk galaxy. The disk extends to at least 1\farcs6
along the major axis, which corresponds to $\sim$8.1~kpc at $z_s=7.455$. Its 
small effective radius (0\farcs16, or 0.81~kpc) is to say that its star light 
distribution is rather concentrated to the central region. 

    From the reconstructed model, we can also get the flux of the object in 
the source plane ($F_{src}$) and that of the arc in the image plane 
($F_{arc}$), which can be easily calculated by summing the pixels. In this 
way, we obtain the lensing magnification of the entire galaxy 
$\mu= F_{arc}/F_{src} = 2.54$. However, the magnification of the dusty region,
$\mu_d$, is different from this value because it is confined within a limited 
region but not spread over the entire galaxy (see Section 4.1). To estimate
$\mu_d$, we take a different approach. As the reconstruction also resulted in
a magnification map, we trace the position of the peak of the ALMA 870~$\mu$m
emission (see Table~\ref{tab:almaherschel}) to this map and take the average 
value of a $3\times 3$ pixel region around this position as $\mu_d$, which is
$\mu_d=3.62\pm0.49$.

    Figure \ref{fig:lensmodel} shows that our lens model also produces a
faint counter image at the position largely coinciding the observed one.
The flux ratio between the arc and its counter-image is $\sim$12.1 in
our model, which is consistent with the measured ratio of $\sim$11.7 (see
Section 4.1).

    We also note that there are three distinct clumps on the arc. After 
deconvolving the observed F444W image by its PSF, we reconstruct the light 
distribution of the source by tracing light rays back to the source plane, and 
the three clumps are identified in the reconstructed source. The corresponding 
positions of the clumps in the observed arc are marked by a black square, 
cross and triangle, respectively. The peak of the ALMA emission lies in
between the cross and the triangle.

   To estimate the halo mass of the lens, we make use of the Einstein radius
$\theta_{\rm E}$, which is related to the redshift of the lens $z_l$, the 
redshift of the source $z_s$ and the velocity dispersion $\sigma_v$ of the 
lens galaxy as
\begin{equation}
\theta_{\rm E} = 4\pi\left(\frac{\sigma_v}{c}\right)^2 \frac{D_{ls}(z_l,z_s)}{D_{s}(z_s)} \ ,
\label{eq:ThetaE}
\end{equation}
where $c$ is the speed of light, $D_{ls}$ and $D_{s}$ are the angular diameter 
distance from the source plane to the lens plane and from the source plane to 
the observer, respectively. Using $z_l = 0.359$, $z_s = 7.455$ and 
$\theta_{\rm E} =$~0\farcs59, we get the velocity dispersion  
$\sigma_v \approx 156$~km~s$^{-1}$. Therefore, the viral mass of the lens is 
$\sim$$2.6\times10^{12} M_{\odot}$.
As shown in Appendix \ref{sedlens}, the stellar mass
of the lens is $\sim$$3.8\times10^{10} M_\odot$. The ratio of the two is
$\sim$68, which is reasonable for a dark-matter-dominated galaxy.

\section{Discussion}

    Considering its $z_{\rm ph}\gtrsim 7$ (and possibly at $z=7.455$; see 
Appendix~\ref{almab3}), the most striking feature of COSBO-7 is its disk 
morphology in the source plane. To maintain such an extended thin disk, the 
galaxy must be fast rotating. Currently, the earliest fast-rotating galaxy
confirmed by spectroscopy is ``Twister-5'' at $z=5.3$ \citep[][]{Nelson2023}. 
If confirmed at $z\gtrsim 7$, COSBO-7 will set a new redshift record for disk 
galaxies and further challenge the existing models where such galaxies always
form late.

    After de-magnifying by $\mu_d=3.62$, the SFR of COSBO-7 in the dusty
region is ${\rm SFR_{IR}} = $~185--232 $M_\odot$~yr$^{-1}$, which makes it a 
starburst regardless 
of the choice of $\beta$ in the MBB fit. As shown in Figure~\ref{fig:mbbfit}, 
it has a very high dust temperature of $\sim$92--126~K (differences due to
different $\beta$), which is even higher than the hot-dust DSFGs 
($\sim$52$\pm$10~K) at $z\approx 1$--2 selected by mid-IR
\citep[][]{Casey2009}. On the other hand, given that the temperature of the
cosmic microwave background is $\sim$23.1~K at $z=7.5$, such a higher dust
temperature at this redshift probably is not very surprising.

    As shown in \citet[][]{Yan2016}, the $L_{\rm FIR}$-T relation of DSFGs
is a manifest of the ``modified'' Stefan-Boltzmann law, based on which the
effective size of the dust emission region (approximated by a sphere) can be 
obtained. From its dust temperature and intrinsic $L_{\rm FIR}$ 
(6.7--8.4$\times 10^{12}L_\odot$), the size of the dusty starburst region 
should be only $\sim$0.15--0.25~kpc in radius based on Figure 2 of 
\citet[][]{Yan2016}.
This is consistent with its being confined to a limited region within the host 
galaxy as argued in Section 4.1. The moderate extinction derived from the 
exposed stellar population (as detected in the JWST images) is also
consistent with a dusty region of a limited size.

    Another notable feature of COSBO-7 is that its exposed stellar population
is unusually bright for an object at $z\gtrsim 7$. Taking $\mu=2.54$ into 
account, its de-magnified brightness is 24.8 to 22.9~mag in the NIRCam bands 
from F200W to F444W (rest-frame $\sim$2300--5060\AA), and it could be 
easily detected without the lensing magnification. Such a brightness in
rest-frame UV-to-optical, contributed by the entire disk, can only be due to 
active, ongoing star formation all over the disk.
Its inferred intrinsic (de-magnified) SFR ranges from 
143~$M_\odot$~yr$^{-1}$ 
to $2.8\times 10^4$~$M_\odot$~yr$^{-1}$.

Even the lowest SFR estimate qualifies it as a starburst, and it is 
possible that it could even be higher than that in the dusty starburst region.
To answer the question posed in Section 1, the exposed stellar population of 
COSBO-7 itself could be a starburst-LBG in the epoch of EoR.

    This exposed stellar population is peculiar also because of its huge
stellar mass as derived. After de-magnifying by $\mu=2.54$, the stellar mass
values reported in Table~\ref{tab:sedfit} correspond 
$M_*=1.4\times 10^{10}$
to $14.0\times 10^{10}M_\odot$.
Even the lowest estimate is 
$>10^{10}M_\odot$, which is very high for galaxies at $z>7$: as the 
age of the universe is only $\sim$700~Myr at $z\approx 7.5$, how could it
have enough time to turn gas into such a huge amount of stars? Nonetheless,
the age estimates are all very small (70--190~Myr), which at least offer a 
self-consistent picture that the starbursting disk formed all its stars 
extremely quickly. 

   Finally, we comment on the dust mass in the dusty starburst region. After
de-magnifying by $\mu_d=3.62$, we get 
$M_d=3.1$, 4.5 and $6.3\times 10^7 M_\odot$ for $\beta=1.5$, 2.0 and 2.5,
respectively, which are very large for $z\approx 7.5$. In order to explain
such a huge amount of dust mass, supernovae must be invoked as the 
contributors \citep[][]{Mancini2015}. Given the high SFR as discussed above, 
it is possible to have a high rate of supernova since its formation to
generate the necessary amount of dust.

    
\section{Summary}

    Using the recent JWST NIRcam and MIRI data, we have studied the true
near-to-mid-IR counterpart of COSBO-7, an SMG known for more than sixteen 
years but not correctly identified in the existing HST images. 
It is in fact a
background, very red galaxy strongly lensed by a foreground galaxy at $z=0.36$
and is only significantly detected at $\lambda>2$~$\mu$m. Fitting its SED
constructed from the NIRCam and MIRI photometry, 
our analysis gives solutions in favor of $z_{\rm ph}>7$. Due to the
statistical nature of the photometric redshift technique, the low-redshift 
possibility cannot be ruled out; however, the high-redshift possibility at
$z>7$ is more likely and cannot be dismissed. The single-line
detection in the ALMA Band 3 data would further put it to $z=7.455$ if the line
is due to the CO (7-6) transition.
Surprisingly, our redshift-independent lens modeling shows 
that it is a thin edge-on disk subtending at least 1\farcs6 in the source 
plane, which corresponds to $\sim$8.1~kpc at $z=7.5$. The galaxy as a whole is 
magnified by a factor of $\mu=2.54$, and the de-magnified brightness in NIRCam 
would still make it one of the brightest F090W-dropouts (candidates at
$z\approx 7$) known to date (22.9~mag in F444W). The inferred intrinsic 
SFR from the four SED fitting tools ranges from of 
143~$M_\odot$~yr$^{-1}$ to 
$2.8\times 10^4$~$M_\odot$~yr$^{-1}$,
which means that the entire stellar disk is experiencing starburst. 

    The dusty star forming region of COSBO-7 is best revealed by the sub-arcsec
ALMA 870~$\mu$m data, which show that the dust emission is on one side of the
galaxy and does not cover the whole stellar disk. The far-IR-to-mm SED 
constructed using the Herschel, ALMA and MOMBO-2 data can be well fitted by a 
single-temperature MBB model, with a high dust temperature of $\sim$92--126~K. 
We argue that this dusty region is confined within a limited region
($\sim$0.15--0.25~kpc in radius) in the galaxy. It is magnified by a larger 
factor of $\mu_d=3.62$, and the intrinsic SFR inferred from its de-magnified 
far-IR luminosity is 185--232~$M_\odot$~yr$^{-1}$. In other words, the dusty
region alone is also experiencing starburst. 

    If it is confirmed at $z>7$ by spectroscopy, all the above will
make COSBO-7 the most exotic galaxy in the EoR. Its stellar mass is
$\sim$$10^{10-11}M_\odot$, and it must have formed nearly all its stars in 
$\sim$70--190~Myr through starburst over the entire galaxy. How it could keep 
its think disk intact is puzzling, especially when considering that it must 
have gone through multiple episodes of intense supernova explosion to generate 
the large amount of dust ($>10^7M_\odot$) necessary to explain the 
far-IR-to-mm emission. To say the least, COSBO-7 will exacerbate the 
challenging situation that the current picture of early galaxy formation has
been facing since the first batch of JWST data were delivered to the community.
Further investigations of this object, especially the kinematics study that
can be enabled by the JWST IFU capabilities, will be critical.

All the JWST data used in this paper can be found in MAST: 
\dataset[10.17909/tw7g-y088]{http://dx.doi.org/10.17909/tw7g-y088}.

\begin{acknowledgements}
We thank the referee for the thoughtful comments, which improve the quality of
this paper.
CL acknowledges the support from the Special Research Assistant program of the 
Chinese Academy of Sciences (CAS). 
HY and BS acknowledge the support from the NSF grant AST-2307447. 
NL acknowledge the support from the science research grants from the China 
Manned Space Project (No. CMS-CSST-2021-A01), the CAS Project for Young 
Scientists in Basic Research (No. YSBR-062), and the Ministry of Science and 
Technology of China (No. 2020SKA0110100).
ZM is supported by the NSF grants No. 1636621, 2034318, and 2307448.
CC is supported by the National Natural Science Foundation of China, 
No. 11933003, 12173045. This work is sponsored (in part) by the CAS, through a
grant to the CAS South America Center for Astronomy (CASSACA). We acknowledge 
the science research grants from the China Manned Space Project with No. CMS-
CSST-2021-A05.

This paper makes use of the following ALMA data: ADS/JAO.ALMA\#2021.1.00705.S, 
ADS/JAO.ALMA\#2016.1.00463.S, and ADS/JAO.ALMA\#2022.1.00863.S. ALMA is a 
partnership of ESO (representing its 
member states), NSF (USA) and NINS (Japan), together with NRC (Canada), MOST 
and ASIAA (Taiwan), and KASI (Republic of Korea), in cooperation with the 
Republic of Chile. The Joint ALMA Observatory is operated by ESO, AUI/NRAO and 
NAOJ. The National Radio Astronomy Observatory (NRAO) is a facility of the 
National Science Foundation operated under cooperative agreement by Associated 
Universities, Inc.
The ALMA data reduction services of this work are supported by the
China-Chile Astronomical Data Center (CCADC), which is  affiliated with the
Chinese Academy of Sciences South America Center for Astronomy 
(CASSACA).

\end{acknowledgements}

\bibliographystyle{aasjournal}

\appendix

\section{Alternative $z_{\rm ph}$ estimate by limiting $z\leq 6$}\label{forcelowz}

    As mentioned in Section 4.2, our SED analysis gives $z_{\rm ph}$ solutions
that are in favor of $z_{\rm ph}>7$ but cannot rule out the low-$z$ 
possibilities due to the statistical nature of photometric redshift. To explore
such low-$z$ possibilities, we repeated the process described in Section 4.2 
but with the difference of setting an upper limit of $z\leq 6$. The results are
summarized in Figure \ref{fig:sedjwst_z6}. 

    Regardless of the F090W upper limit choice, LePhare gives solutions close to
the upper end of the allowed redshift range, while the other three result in 
solutions at $z_{\rm ph} \approx 2$--4. However, all these solutions have
However, all these solutions have much worsemuch worse $\chi^2$ than those presented in Section 4.2: in case of
using the F090 limit of 28.43~mag, the difference of the $\chi^2$ values 
range from $\Delta\chi^2 \approx 15$--65; in case of using the limit of
26.98~mag, $\Delta\chi^2 \approx 6$--9. Therefore, the high-$z$ solutions in
Section 4.2 are indeed preferred.

\setcounter{figure}{0}
\renewcommand{\thefigure}{A\arabic{figure}}

\begin{figure*}
    \centering
    \includegraphics[width=0.9\textwidth,height=\textheight,keepaspectratio]{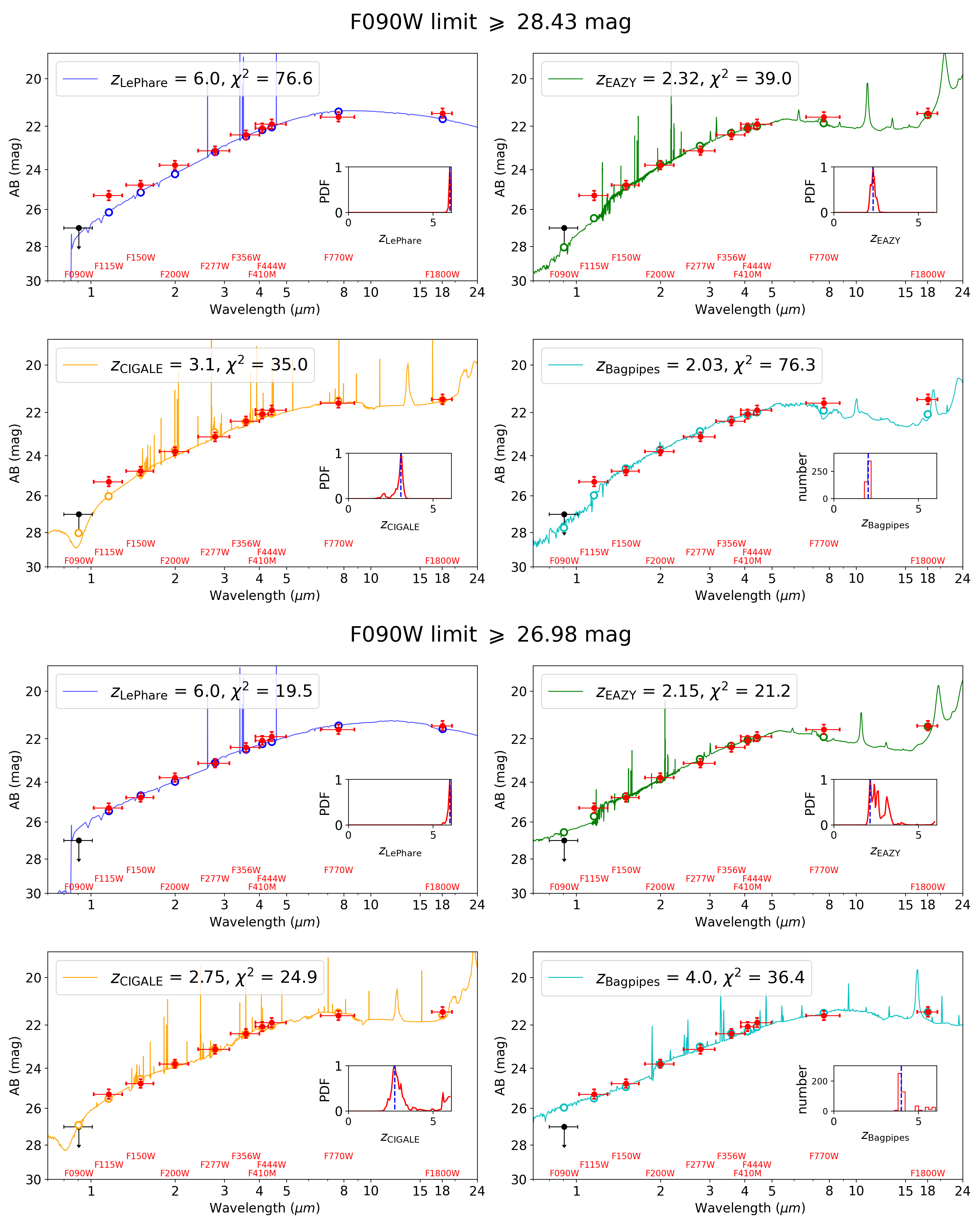}
    \caption{
    Same as Figure~\ref{fig:sedjwst} but with the limit of $z\leq 6$ when
    running the fitting codes to obtain low-$z$ solutions.}
    
    \label{fig:sedjwst_z6}

\end{figure*}

    For the sake of completeness, we also performed the MBB fit of the 
far-IR-to-mm SED as in Section 4.3 but at the fixed redshift of $z=4$. Due to
the degeneracy of dust temperature and redshift, the high temperature obtained
in Section 4.3 is lowered in this case because of the adopted lower redshift.

\begin{figure*}
    \centering
    \includegraphics[width=\textwidth,height=\textheight,keepaspectratio]{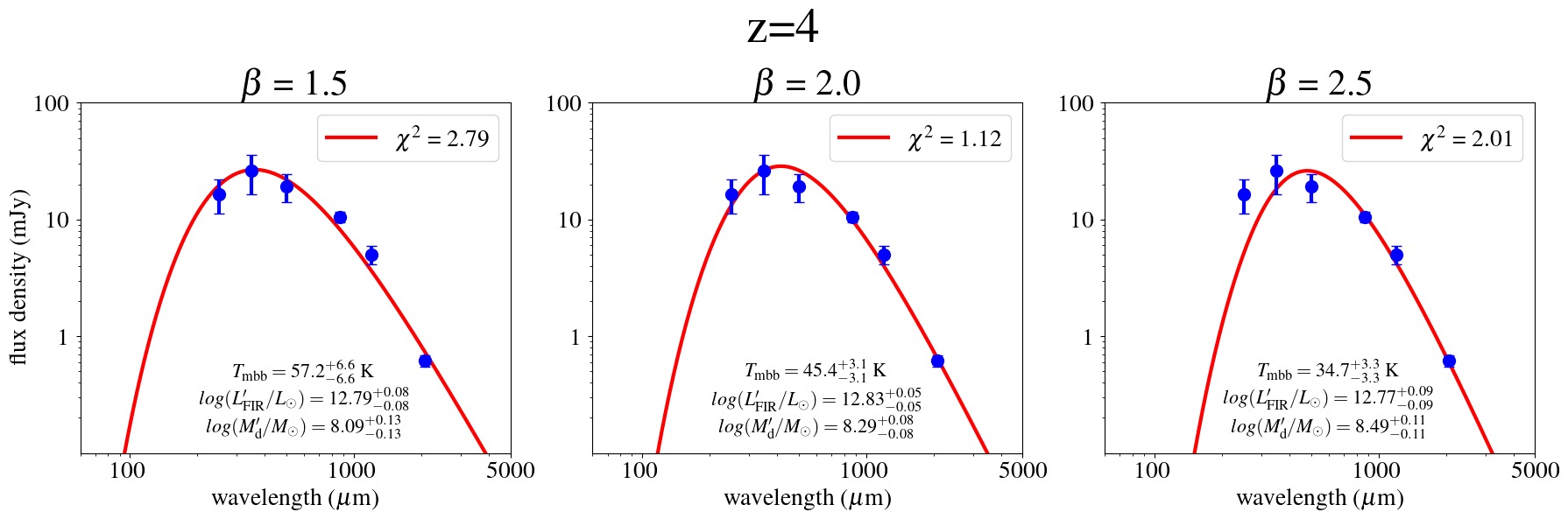}
    \caption{
    Same as Figure~\ref{fig:mbbfit} but done at $z=4$.}
    
    \label{fig:mbbfit}
\end{figure*}

\setcounter{figure}{0}
\renewcommand{\thefigure}{B\arabic{figure}}

\begin{figure}
    \centering
    \includegraphics[width = 0.5\textwidth]{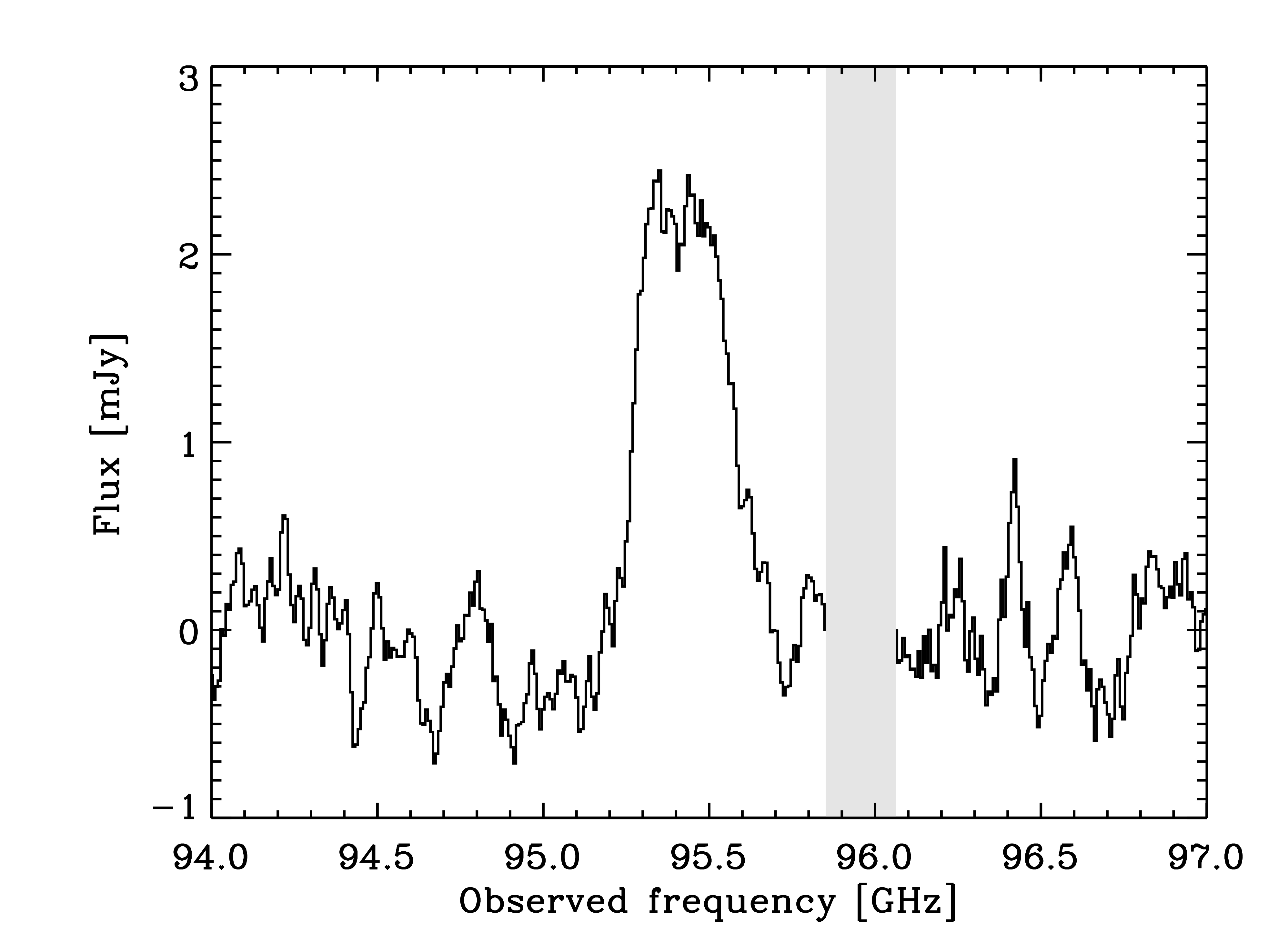}
    \caption{
    ALMA Band 3 spectrum of COSBO-7 that reveals the line of double-peak
feature. If the line is due to the CO (7-6) transition (at the rest-frame
806.65~GHz), the corresponding redshift is $z=7.455$ and is consistent with the
$z_{\rm ph}$ determined in Section 4.2. The grey region represents the frequency 
ranges not covered by the observations.
}
    \label{fig:band3line}
\end{figure}

\section{ALMA Band 3 Line Detection}\label{almab3}

   COSBO-7 was observed in ALMA Band 3 in January to March 2023 over the range
of 84--107~GHz in three separate sectional scans (PID 2022.1.00863.S, 
PI: J. Hodge), and the data of interest were released while this manuscript was
under revision. We reduced these data using the default pipeline 
{\sc ScriptForPi.py} to obtain the visibility file and performed {\sc tclean}
using CASA. A line was clearly seen.
The continuum was determined from the channels free of line 
emissions and removed in the {\it uv} space by {\sc uvcontsub}. We adopted the 
natural weighting parameters for the further process. We set the pixel scale to 
0\farcs25 and the channel width to 7.812~MHz, which is the same as the original 
resolution of the observations. The final data cube has the beam of
(bmaj, bmin, PA) = (1\farcs60, 1\farcs43, $-$60.35$^{\rm o}$) and an rms of 
0.317~mJy~beam$^{-1}$.

   We extracted the spectrum using the CASA task {\sc viewer}. The extraction
was done at the Band 4 image position (see Table~\ref{tab:almaherschel}) with a 
circular aperture of 4\arcsec\ in diameter. The result is shown in
Figure~\ref{fig:band3line}. The line is centered at 95.4~GHz, with a clear
double-peak feature. If the line is attributed to the CO (7-6) transition at
the rest-frame 806.65~GHz, the corresponding redshift is $z=7.455$, which is
consistent with the $z_{\rm ph}$ derivation in Section 4.2. However, such a
single line is not sufficient to unambiguously determine the redshift because
it could also be due to other lines such as CO (6-5) (corresponding to 
$z=6.248$), CO (5-4) (corresponding to $z=5.04$), CO (4-3) (corresponding to 
$z=3.833$), etc. Detecting other lines will be necessary to nail down its
redshift.

   We also note that the double-peak feature indicates a relative velocity of
$\sim$270~km~s$^{-1}$ if the line is the CO (7-6) transition. As we argued in 
Section 4.1, the dusty starburst region is likely confined within a small
region, and therefore this velocity should not be interpreted as the rotation
velocity of the disk. In other words, it is more likely due to two merging
components moving along the sightline.

  Lastly, we comment on the non-detection of the [C~I] line (rest-frame 
809.34~GHz), which is often seen accompanying the CO (7-6) line. The ratio 
of these two lines ([C~I]/CO(7-6), in logarithmic scale), however, depends on 
temperature, and the [C~I] line could vanish at high temperatures. Using the 
ratio of the flux densities at 60 and 100~$\mu$m (C(60/100)) as the proxy to 
temperature, it was shown in \citet{Lu2017, Lu2018} that [C~I]/CO(7-6) 
decreases with C(60/100). By coincidence, at $z=7.455$ the rest-frame 60 and
100~$\mu$m correspond to approximately Herschel SPIRE 500~$\mu$m and ALMA 
870~$\mu$m, respectively. From Table~\ref{tab:almaherschel}, the ratio of the 
two is $\sim$1.8. Based on the relation given in \citet[][]{Lu2018}, the [C~I]
line should be $\sim$11.5$\times$ weaker than the CO (7-6) line, and the 
sensitivity of the current data is not sufficient for its detection.

\section{SED fitting of the foreground lens}\label{sedlens}

   The catalog of \citet[][]{Laigle2016} includes the stellar mass estimate of
the foreground galaxy at $z=0.359$, which is 
${\rm log}(M_*/M_{\odot})=10.66^{+0.03}_{-0.04}$. As this quantity provides
some constraints on the validity of this galaxy being a viable lens, we
have derived it again by fitting its optical-to-near-IR SED after incorporating
the NIRCam photometry reported in Table~\ref{tab:jwstphot} and fixing its
redshift at $z=0.36$. The HST photometry
is taken from \citet[][]{Nayyeri2017}, and the magnitudes are $21.167\pm 0.008$
and $20.157\pm 0.004$ in the ACS F606W and F814W and $19.407\pm 0.003$ and 
$19.095\pm 0.002$~mag in the WFC3-IR F125W and F160W, respectively. The results
are summarized in Figure~\ref{fig:sedlens}. These stellar mass estimates are
consistent with the result from \citet[][]{Laigle2016}. We take the average
of our results, which is $3.8\times10^{10}M_\odot$.

\setcounter{figure}{0}
\renewcommand{\thefigure}{C\arabic{figure}}

\begin{figure*}
    \centering
    \includegraphics[width=\textwidth,height=\textheight,keepaspectratio]{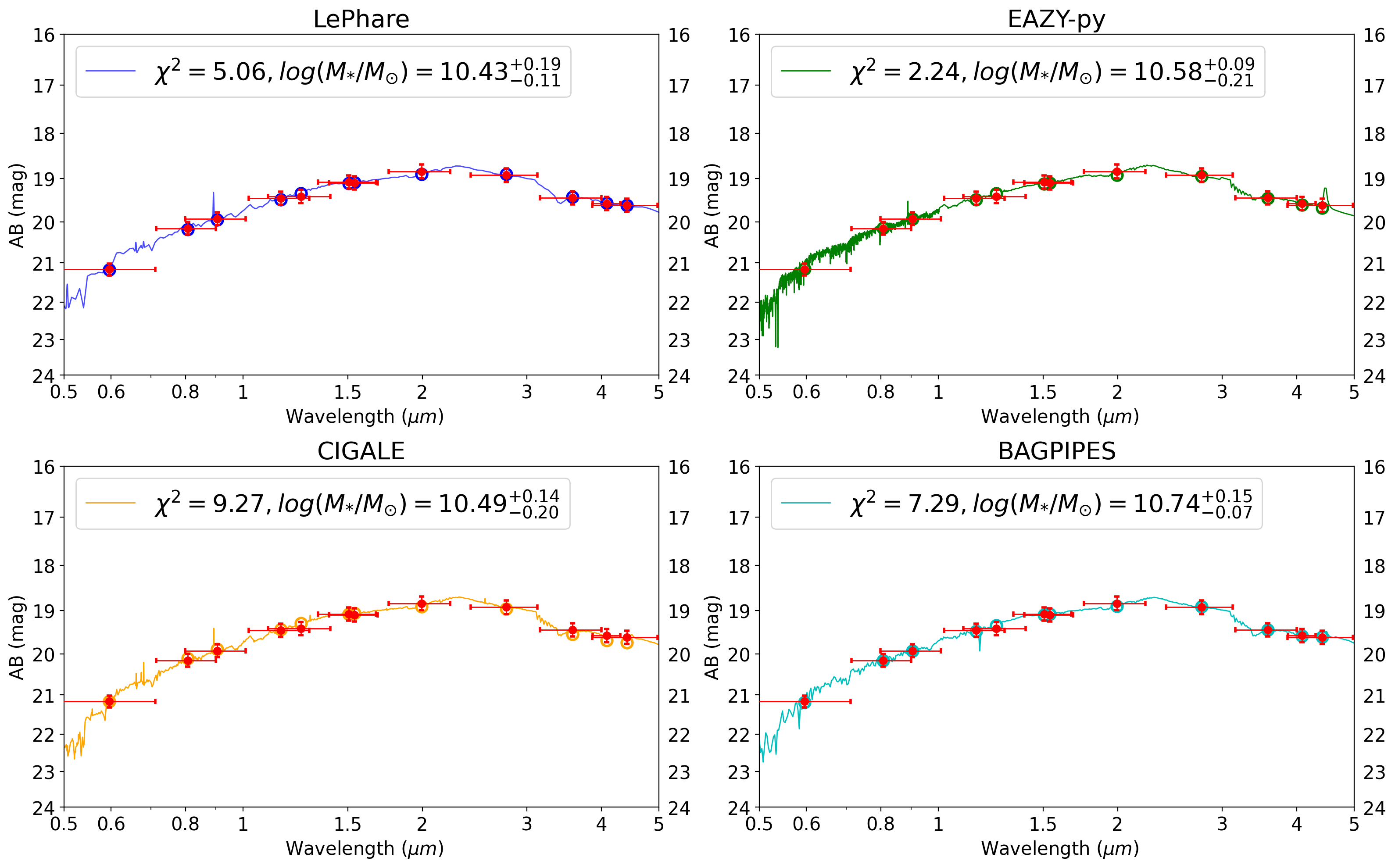}
    \caption{SED fitting results for the foreground lens galaxy at $z=0.36$,
    obtained following the same procedures as in Figure~\ref{fig:sedjwst}.} 
    \label{fig:sedlens}
\end{figure*}

\end{document}